# PROMPT PHOTON PRODUCTION IN A BREMSSTRAHLUNG $qq \to qq\gamma$ IN PROTON-PROTON COLLISIONS AT $\sqrt{s}$=10 GeV NICA ENERGIES

**M.R. Alizada**[1], **A.I.Ahmadov**[1,2]

[1]Department of Theoretical Physics, Baku State University

Z.Khalilov 33, Az-1148 Baku, Azerbaijan, e-mail: mohsunalizade@gmail.com

[2]Institute for Physical Problems, Baku State University, Baku, Azerbaijan

Z.Khalilov 33, Az-1148 Baku, Azerbaijan

## ABSTRACT

The dependence of differential cross-section of prompt photon production in bremsstrahlung $qq \to qq\gamma$ at collisions of nonpolarized and longitudinally polarized protons at $\sqrt{s}$=10 GeV NICA energies on the kinematic parameters: sum of energy of the colliding protons $\sqrt{s}$, the transverse momentum $p_T$, the cosine of the scattering angle $Cos(\theta)$, the rapidity $y$ of photon and $x_T$ has been investigated.

Differential cross-section of bremsstrahlung accounts for 0.03% of the total diferential cross section for the production of prompt photons in a proton-proton collisions at NICA entrgies. The polarization of protons has a large influence on the differential cross-section of bremsstrahlung $qq \to qq\gamma$, at large values of the transverse momentum of prompt photons.

Double spin asymmetry of process of bremsstrahlung on the kinematic parameters has been studied.

## I. INTRODUCTION

Prompt photons are photons that are produced directly by the hard scattering of partons such as quarks and gluons before they hadronize into mesons and baryons. Photons produced in proton–proton collisions with energies from 1.5 GeV to several GeV carry information about the formation of the parton distribution in nucleons, since they are produced as a result of hard parton scattering, such as the Compton scattering of the quark–gluon $qg\rightarrow q\gamma$ and the annihilation of the quark–antiquark pair $q\bar{q}\rightarrow g\gamma$, as well as the bremsstrahlung of the $qq\rightarrow qq\gamma$ quarks and the quark–gluon phase. Prompt photon production plays a significant role in determining the gluon distribution in the proton and testing aspects of perturbative quantum chromodynamics (pQCD). The Spin Physics Detector (SPD) experiment, which focuses on spin physics, also benefits significantly from studying direct photons, as they can provide valuable information about spin-dependent parton distributions. Studying the spectra of prompt photons produced in proton–proton collisions has attracted considerable attention in high-energy physics. This process plays a key role in understanding the mechanisms of particle interactions and the structure of protons. Analysis of the subprocesses involving sea quarks and gluons allows for a more detailed study of the proton structure [1-3].

There are many theoretical and experimental investigation devoted to various aspects of prompt photon production in proton-proton collisions at the LHC and American Tevatron energies [4-6]. The theoretical calculations include pQCD, which takes into account the strong interactions at short distances where the strength coupling constant $\alpha_s$ is weak enough for a perturbative approach. The calculations also include parton distribution function (PDFs), which describe the probability of finding a parton with a given momentum fraction inside a proton, and fragmentation functions (FFs), which describe how partons transform into hadrons. The measurements cover a wide range of transverse momentum and pseudo-design photons and are compared with various theoretical predictions, including QCD calculations in the next-to leading order (NLO) and Monte Carlo (MC) events generators.

Studies of the production of prompt photons in proton-proton collisions at the energies of the LHC and American Tevatron have shown that the subprocesses of Compton scattering quark-gluon $qg \to q\gamma$ and annihilation of a quark-antiquark pair $q\bar{q} \to g\gamma$ are the dominant processes of production of promptt photons, and their cross sections make up approximately 95% and 5% of the total cross-section of produced prompt photons [4-6].

Bremsstrahlung is one of main subprocesses of production of prompt photon in proton-proton collisions. Different methods were applied to studying the process of production of prompt photons in bremsstrahlung, such as: Hydrodynamic, pQCD, logarithmic approximation (harmonic oscillator), etc. [7-14]. The differential sections of this process, its deposits, the spectrum of energy of photons of brake radiation, and loss of energy with quarks, etc. were investigated.

In the paper [7], is considered the problem of medium effects on single gluon bremsstrahlung associated with producing a high-energy particle in a finite, time-dependent QCD plasma. Working to leading logarithmic order, the author shows that the result for the bremsstrahlung gluon spectrum can be cast into a remarkably simple form in the general case. Similarly has been analyzed the process of pair production and comments on the radius of convergence of the opacity expansion in cases where the leading-log approximation holds, showing that the opacity expansion does not converge when the thickness of the plasma is greater than roughly the bremsstrahlung formation time. The effect of critical temperature, the communication constant and energy of photons on the speed of photon production are investigated. The work found that the use of a high critical temperature allows you to increase the force of the connection of the interaction of quarks and its sensitivity to the speed of photons. The work also discusses the influence of energy, in particular, energies in the range (from 1.5 to 5 GeV) for the speed of photons.

In the paper [8], Fiol B. and his co-authors propose an exact formula for the energy radiated by an accelerating quark in N=2 superconformal theories in four dimensions. This formula, expressed in terms of the so-called bremsstrahlung function B, reproduces the known result for $N$=4 theories and provides a prediction for all perturbative and instanton corrections in $N$=2 theories. The authors perform a perturbative check of their proposal up to three loops.

The paper [9] analyzes photon production from a nonequilibrium QGP. The authors derive an integral equation that describes photon production through quark-antiquark annihilation and quark bremsstrahlung. The equation includes coherence between different scattering sites, also known as the Landau-Pomeranchuk-Migdal (LPM) effect. These leading-order (LO) processes are studied for the first time together in an out-of-equilibrium field-theoretical treatment that enables the inclusion of viscous corrections to the calculation of electromagnetic emission rates. In the special case of an isotropic, viscous plasma, the integral equation depends on only three constants, which capture the nonequilibrium nature of the medium.

In the paper [10], Song T. calculates the bremsstrahlung photon emission originating from the hadronization process of a QGP. Assuming that quark and antiquark numbers are conserved during hadronization, the author obtains their densities from a statistical hadron model and uses these densities to calculate bremsstrahlung in the soft-photon approximation. It is found that this contribution to the direct photon spectrum becomes more significant in low-energy heavy-ion collisions and in peripheral collisions, where the QGP lifetime is relatively short, making it a potentially important source of direct photons, especially at the late stage of the system's evolution.

Studies of Bayer, Dokshitser, Muller and Schiff [11-13] gave a direct docking of how the environment affects the probability of radiation of one glue in the inhibitory method with a high-energy parton.

The paper [11] studies the medium-induced energy loss spectrum of a high energy $E$ quark or gluon passing through a hot QCD medium of finite volume. The interaction is modeled by a simple picture of static scattering centers. It is found that the total induced energy loss grows as $L^2$, where $L$ is the extent of the medium. This quadratic behavior is true for $L$ smaller than the critical length $L_{cr} \propto \sqrt{E}$. Solving the energy loss problem reduces to solving a Schrödinger-type equation whose "potential" is given by the cross section of a single scattering of a high-energy parton in the medium. These results should be directly applicable to QGP.

The paper [12] analytically studies the medium-induced energy loss of a high-energy parton passing through a finite-sized QCD plasma that expands longitudinally according to the Björken model. The authors extend the Bayer-Dokshitzer-Muller-Penier-Schiff (BDMPS) formalism, previously applied to static media, to the case where a quark passes through successive layers of matter with decreasing temperature. It is shown that the resulting radiative energy losses can be 6 times larger than in a static plasma at a finite temperature $T = T(L)$, which is reached after the quark has traveled a distance $L$.

In the paper [13], the authors extend the BDMPS formalism for calculating radiative energy loss to the case where the emitted gluon carries away a finite fraction of the quark momentum. Some previously omitted virtual corrections are included in the calculations. The equivalence of the BDMPS formalism and the Zakharov B. formalism is clearly demonstrated.

A high-energy parton can be formed by some complicated process against the background of a homogeneous, time-invariant part of hot QCD matter. In subsequent works by Bayer, Dokshitzer, Müller and Schiff, using an expanding QGP as an example, the authors showed that logarithmic results can also be obtained for inhomogeneous, time-dependent media, But the result was not so simple, since it required double integration of the complicated function that was found for the specific example they considered.

The calculation for purely hadronic large $p_T$ scattering is very difficult due to the much larger number of subprocesses involved. Research into the production of prompt photons in proton-proton collisions at Nuclotron-based Ion Collider Facility (NICA) 10-27 GeV energies has its advantages. The energy of NICA does not allow the production of many elementary particles, which makes it difficult to accurately determine the differential cross-section of prompt photons production. Moreover, at NICA energies is possible transition QGP to adronization. From this point of view, the studies that are planned to be carried out at the NICA experimental facility are important [14]. At the NICA energy are planed investigation with polarized protons

[15].

The articles [16,17] are discussed by the possible formation of the QGP in the NICA energies. Ivanishchev D.A. and co-authors investigate the feasibility of studying the properties of thermal photons in heavy ion collisions at the NICA accelerator, which is currently under construction. Thermal photons are a crucial source of information about the properties of hot and dense matter - QGP formed in such collisions. The effective temperature of the medium, measured using thermal photons at RHIC and LHC, significantly exceeds the expected QGP phase transition temperature. Studying prompt photons at NICA can help estimate the effective temperature of the medium at lower energies and trace the transition from partonic to hadronic degrees of freedom. The paper presents the results of a study on the feasibility of measuring thermal photons in the Multi-Purpose Detector (MPD) experiment at NICA in gold-gold (Au-Au) collisions using the method of photon conversion into electron-positron pairs.

Our early research of prompt photon production in nonpolarized and logitudinally polarized proton-proton collisions in LO and NLO approximation at NICA energies showed that differential cross section of prompt photons production consists of more than 50% of the differential cross section of the subprocess of Compton scattering of quark-gluon and about 43% of the differential cross section of annihilation of quark-antiquark pairs [18-20]. Despite the low collisions NICA energies, Compton scattering of quark-gluon $qg \to g\gamma$ remains the dominant subprocess as at LHC energies.

In this article we present the results of investigation of prompt photon production in bremsstrahlung $qq \to qq\gamma$ in collisions of nonpolarized, and longitudinally polarized protons at NICA energies.

## II. DIFFERENTIAL CROSS-SECTION OF THE BREMSSTRAHLUNG $qq \to qq\gamma$ OF PROMPT PHOTON PRODUCTION

The prompt photons production in bremsstrahlung $qq \to qq\gamma$ is described with 16 Feynman diagrams, which are generated by FeynArts [21]. Matrix elements of process are created by FeynCalc [22]. Feynman diagrams and matrix elements of process are shown in Appendix.

### *2.1. Differential cross section of the bremsstrahlung subprocess $qq \to qq\gamma$ of prompt photon production in collisions of nonpolarized protons*

The square of the matrix element, averaged over the spin of the initial particles, is:

$$|\bar{M}|^2 = -\frac{\alpha_e e_{q_i} e_{q_j}\left(\hat{q}_1^2(\hat{s}+2\hat{u}) - \hat{q}_1\left(2\hat{q}_2(\hat{s}+\hat{t}+\hat{u}) + \hat{u}(\hat{u}-\hat{t}) + \hat{q}_2\left(\hat{q}_2(\hat{s}+2\hat{t}) + \hat{t}(\hat{u}-\hat{t})\right)\right)\right)}{39366\pi\hat{q}_1\hat{q}_2\hat{t}\hat{u}(\hat{q}_1-\hat{q}_2+\hat{t})(\hat{q}_1-\hat{q}_2-\hat{u})(\hat{q}_1+\hat{s}+\hat{t})(\hat{q}_1+\hat{s}+\hat{u})}$$

$$\times \Big(2048\pi^4\alpha_e^2\big(3\hat{q}_1^3(2\hat{q}_2+\hat{s}+\hat{u}) + 9\hat{q}_1^2(\hat{s}+\hat{t})(\hat{q}_2+\hat{s}+\hat{u}) + \hat{q}_1\big(6\hat{q}_2^3 + 9\hat{q}_2^2(\hat{s}+\hat{u}) + \hat{q}_2(26\hat{s}^2 + 26\hat{s}(\hat{t}+\hat{u})$$

$$+13\hat{t}^2 + 8\hat{t}\hat{u} + 13\hat{u}^2)\big) + (\hat{s}+\hat{u})(16\hat{s}^2 + 22\hat{s}\hat{t} + 10\hat{s}\hat{u} + 11\hat{t}^2 + 4\hat{t}\hat{u} + 5\hat{u}^2)\big)$$

$$+ 3\hat{q}_2^3(\hat{s}+\hat{t}) + 9\hat{q}_2^2(\hat{s}+\hat{t})(\hat{s}+\hat{u}) +$$

$$+\hat{q}_2(\hat{s}+\hat{t})(16\hat{s}^2 + 10\hat{s}\hat{t} + 22\hat{s}\hat{u} + 5\hat{t}^2 + 4\hat{t}\hat{u} + 11\hat{u}^2) + 6\hat{s}(\hat{s}+\hat{t})(2\hat{s}^2 + 2\hat{s}\hat{t} + \hat{t}^2)$$

$$+ \hat{u}(24\hat{s}^3 + 40\hat{s}^2\hat{t} + 22\hat{s}\hat{t}^2 + 5\hat{t}^3)$$

$$+\hat{u}^3(6\hat{s}+5\hat{t}) + 2\hat{u}^2(\hat{s}+\hat{t})(9\hat{s}+2\hat{t})\big)$$

$$- 768\pi^2\alpha_e\alpha_s(2\hat{s}^2 + 2\hat{s}(\hat{t}+\hat{u}) + (\hat{t}+\hat{u})^2)(\hat{q}_1(2q_2+\hat{s}+\hat{u}) + \hat{q}_2(\hat{s}+\hat{t}) + \hat{t}\hat{u})$$

$$+9\alpha_s^2\big(3\hat{q}_1^3(2\hat{q}_2+\hat{s}+\hat{u}) + 9q_1^2(\hat{s}+\hat{t})(\hat{q}_2+\hat{s}+\hat{u})$$

$$+ \hat{q}_1\big(6\hat{q}_2^3 + 9\hat{q}_2^2(\hat{s}+\hat{u}) + \hat{q}_2(34\hat{s}^2 + 34\hat{s}(\hat{t}+\hat{u}) + 17\hat{t}^2 + 16\hat{t}\hat{u} + 17\hat{u}^2)$$

$$+(\hat{s}+\hat{u})(20\hat{s}^2 + 26\hat{s}\hat{t} + 14\hat{s}\hat{u} + 13\hat{t}^2 + 8\hat{t}\hat{u} + 7\hat{u}^2)\big) + 3\hat{q}_2^3(\hat{s}+\hat{t}) + 9\hat{q}_2^2(\hat{s}+\hat{t})(\hat{s}+\hat{u})$$

$$+ \hat{q}_2(\hat{s}+\hat{t})(20\hat{s}^2 + 14\hat{s}\hat{t} + 26\hat{s}\hat{u} + 7\hat{t}^2$$

$$+8\hat{t}\hat{u} + 13\hat{u}^2) + 6\hat{s}(\hat{s}+\hat{t})(2\hat{s}^2 + 2\hat{s}\hat{t} + \hat{t}^2) +$$

$$\hat{u}(24\hat{s}^3 + 44\hat{s}^2\hat{t} + 22\hat{s}\hat{t}^2 + 7\hat{t}^3) + \hat{u}^3(6\hat{s}+7\hat{t}) + 2\hat{u}^2(\hat{s}+\hat{t})(9\hat{s}+4\hat{t})\big)\Big).$$

The differential cross section at the parton level is determined by following expression:

$$d\hat{\sigma} = \frac{1}{2\hat{s}} \mid \bar{M} \mid^2 \; d\Phi_3,$$

where $\hat{s} = (p_1 + p_2)^2$, and $d\Phi_3$ is the three-body Lorentz-invariant phase space. The Lorentz-invariant three-particle phase space is defined as

$$d\Phi_3 = (2\pi)^4 \delta^{(4)}(p_1 + p_2 - p_3 - p_4 - k) \prod_{i=3,4,\gamma} \frac{d^3 p_i}{(2\pi)^3\, 2E_i}.$$

Using standard factorization of the three-body phase space:

$$d\Phi_3(P; p_3, p_4, k_1) = d\Phi_2(P; Q, k_1) \frac{d\hat{s}_1}{2\pi} d\Phi_2(Q; p_3, p_4),$$

where: $P = p_1 + p_2,\ Q = p_3 + p_4$ .

For massless particles ($p_i^2 = 0$): $d\Phi_2(P; q, k) = \frac{1}{8\pi} \frac{|\vec{k}|}{\sqrt{\hat{s}}} d\Omega_\gamma$, $d\Phi_3 = \frac{1}{256\pi^3} \frac{1}{\hat{s}}\, d\hat{s}_1\, d\hat{t}$.

After integrating over trivial angles, one obtains:

$$\frac{d^2\hat{\sigma}}{d\hat{s}_1 d\hat{t}} = \frac{1}{512\pi^3 \hat{s}^2} \mid \bar{M} \mid^2.$$

The invariant $\hat{s}_1$ varies within the limits: $0 \leq \hat{s}_1 \leq \hat{s}$.

The differential cross section at the hadron level is determined as:

$$d\sigma = G_{q_1/h_1}(x_1) G_{q_2/h_2}(x_2) dx_1 dx_2 \hat{\sigma}, \tag{1}$$

where $\hat{\sigma} = \int \frac{d\hat{\sigma}}{d\hat{t}} d\hat{t}$ – cross-section of subprocess at the parton level, $G_{q_1/h_1}(x_1)$ and $G_{q_2/h_2}(x_2)$ - parton distribution function, [23], $x_1$ and $x_2$ fractions of the momentum carried away by partons from the total momentum of protons.

It is also necessary to indicate, the relation between the Mandelstam invariants for the subprocess - parton level through the Mandelstam invariants of the main process $pp \to \gamma X$:

$$\hat{s} = x_1 x_2 s, \quad \hat{t} = x_1 t, \quad \hat{u} = x_2 u, \quad t = -p_T\sqrt{s}e^{-y}, \quad u = -p_T\sqrt{s}e^{y}, \tag{2}$$

where $\hat{s}$, $\hat{t}$ и $\hat{u}$ − Mandelstam invariants at the parton level, $s$, $t$ и $u$ − Mandelstam invariants at the hadronic level, where, $\sqrt{s}$ − sum of energy of colliding protons, $p_T$ − transverse momentum of prompt photon and $y$ − rapidity of prompt photon.

Transition from $d\hat{\sigma}/dt$ to $d\sigma/dy$ and $d\sigma/dp_T^2$, we need to use the Jacobian of the transition from dt to y and $p_T$, which have the following form:

$$dt = -p_T\sqrt{s}e^{-y}dy = -tdy, \tag{3}$$

$$dt = -\sqrt{s}e^{-y}dp_T = \frac{|t|}{2p_T^2}dp_T^2, \tag{4}$$

where $dp_T^2 = 2p_T dp_T$.

Substituting last obtaining expressions (2) and (3) into expression (1), we obtain:

$$d\sigma = \int(-t)G_{q_1/h_1}(x_1)G_{q_2/h_2}(x_2)dx_1dx_2\frac{d\hat{\sigma}}{d\hat{t}}dy, \tag{5}$$

$$\frac{d\sigma}{dy} = \int(-t)G_{q_1/h_1}(x_1)G_{q_2/h_2}(x_2)dx_1dx_2\frac{d\hat{\sigma}}{d\hat{t}}. \tag{6}$$

Substituting expression (2) into expression (1) we obtain:

$$\frac{d\sigma}{dp_T^2} = \int(\frac{-t}{2p_T^2})G_{q_1/h_1}(x_1)G_{q_2/h_2}(x_2)dx_1dx_2\frac{d\hat{\sigma}}{d\hat{t}} \tag{7}$$

Similarly obtained $d\sigma/dx_T^2$ and $d\sigma/dCos(\theta)$.

$$\frac{d\sigma}{dx_T^2} = \int dx_1dx_2\, G_{q_1/h_1}(x_1)\, G_{q_2/h_2}(x_2)\, \frac{d\hat{\sigma}}{d\hat{t}}\, \frac{\hat{s}}{4|\mathrm{Cos}(\theta)^*|} \tag{8}$$

$$\frac{d\sigma}{d\mathrm{Cos}(\theta)} = \int dx_1 dx_2\, G_{q_1/h_1}(x_1)\, G_{q_2/h_2}(x_2)\, \frac{d\hat{\sigma}}{d\hat{t}}\, \frac{\hat{s}}{2} \tag{9}$$

To ensure consistency in the evaluation of PDFs under scale variation, we used LHAPDF version 6.5.5 to access modern PDF sets with built-in support for scale-dependent evolution [24]. This allows us to study how the variation in scale impacts both the normalization and shape of differential distributions.

### *2.2 Differential cross-section of the bremsstrahlung subprocess $qq \to qq\gamma$ of prompt photon production in collisions of longitudinally polarized protons*

The Feynman diagrams, matrix elements and Mandelstam invariants of the bremsstrahlung subprocess $qq \to qq\gamma$ taking into account the longitudinal polarization of the colliding protons have a similar appearance as for the case collisions of nonpolarized protons.

The longitudinal polarization of colliding protons is taken into account as in [25].

The following formula is obtained for the square of the modulus of the matrix element of the bremsstrahlung subprocess $qq \to qq\gamma$, taking into account the spin averaging over the final particles and the longitudinal polarization of the interacting particles:

$$|\bar{M}|^2_{POL} =$$

$$= -\frac{\alpha_e e_{q_i} e_{q_j} (1+\lambda_1\lambda_2)\left(\hat{q}_1^2(\hat{s}+2\hat{u}) - \hat{q}_1\left(2\hat{q}_2(\hat{s}+\hat{t}+\hat{u}) + \hat{u}(\hat{u}-\hat{t}) + \hat{q}_2\left(\hat{q}_2(\hat{s}+2\hat{t}) + \hat{t}(\hat{u}-\hat{t})\right)\right)\right)}{39366\pi\hat{q}_1\hat{q}_2\hat{t}\hat{u}(\hat{q}_1-\hat{q}_2+\hat{t})(\hat{q}_1-\hat{q}_2-\hat{u})(\hat{q}_1+\hat{s}+\hat{t})(\hat{q}_1+\hat{s}+\hat{u})}$$

$$\times (2048\pi^4\alpha_e^2(3\hat{q}_1^3(2\hat{q}_2+\hat{s}+\hat{u}) + 9\hat{q}_1^2(\hat{s}+\hat{t})(\hat{q}_2+\hat{s}+\hat{u})$$

$$+ \hat{q}_1(6\hat{q}_2^3 + 9\hat{q}_2^2(\hat{s}+\hat{u}) + \hat{q}_2(26\hat{s}^2 + 26\hat{s}(\hat{t}+\hat{u}) +$$

$$+13\hat{t}^2 + 8\hat{t}\hat{u} + 13\hat{u}^2)) + (\hat{s}+\hat{u})(16\hat{s}^2 + 22\hat{s}\hat{t} + 10\hat{s}\hat{u} + 11\hat{t}^2 + 4\hat{t}\hat{u} + 5\hat{u}^2)\Big) + 3\hat{q}_2^3(\hat{s}+\hat{t})$$

$$+ 9\hat{q}_2^2(\hat{s}+\hat{t})(\hat{s}+\hat{u}) +$$

$$+\hat{q}_2(\hat{s}+\hat{t})(16\hat{s}^2 + 10\hat{s}\hat{t} + 22\hat{s}\hat{u} + 5\hat{t}^2 + 4\hat{t}\hat{u} + 11\hat{u}^2) + 6\hat{s}(\hat{s}+\hat{t})(2\hat{s}^2 + 2\hat{s}\hat{t} + \hat{t}^2)$$

$$+ \hat{u}(24\hat{s}^3 + 40\hat{s}^2\hat{t} + 22\hat{s}\hat{t}^2 + 5\hat{t}^3) +$$

$$+\hat{u}^3(6\hat{s}+5\hat{t}) + 2\hat{u}^2(\hat{s}+\hat{t})(9\hat{s}+2\hat{t})\Big)$$

$$- 768\pi^2\alpha_e\alpha_s(2\hat{s}^2 + 2\hat{s}(\hat{t}+\hat{u}) + (\hat{t}+\hat{u})^2)(\hat{q}_1(2q_2+\hat{s}+\hat{u}) + \hat{q}_2(\hat{s}+\hat{t}) + \hat{t}\hat{u}) +$$

$$+9\alpha_s^2(3\hat{q}_1^3(2\hat{q}_2+\hat{s}+\hat{u})+9q_1^2(\hat{s}+\hat{t})(\hat{q}_2+\hat{s}+\hat{u})$$
$$+\hat{q}_1(6\hat{q}_2^3+9\hat{q}_2^2(\hat{s}+\hat{u})+\hat{q}_2(34\hat{s}^2+34\hat{s}(\hat{t}+\hat{u})+17\hat{t}^2+16\hat{t}\hat{u}+17\hat{u}^2)+$$
$$+(\hat{s}+\hat{u})(20\hat{s}^2+26\hat{s}\hat{t}+14\hat{s}\hat{u}+13\hat{t}^2+8\hat{t}\hat{u}+7\hat{u}^2))+3\hat{q}_2^3(\hat{s}+\hat{t})+9\hat{q}_2^2(\hat{s}+\hat{t})(\hat{s}+\hat{u})$$
$$+\hat{q}_2(\hat{s}+\hat{t})(20\hat{s}^2+14\hat{s}\hat{t}+26\hat{s}\hat{u}+7\hat{t}^2+$$
$$+8\hat{t}\hat{u}+13\hat{u}^2)+6\hat{s}(\hat{s}+\hat{t})(2\hat{s}^2+2\hat{s}\hat{t}+\hat{t}^2)+\hat{u}(24\hat{s}^3+44\hat{s}^2\hat{t}+22\hat{s}\hat{t}^2+7\hat{t}^3)+\hat{u}^3(6\hat{s}+$$
$$7\hat{t})+2\hat{u}^2(\hat{s}+\hat{t})(9\hat{s}+4\hat{t}))\Big).$$

The differential cross-section at the parton level is determined by the expression:

$$\frac{d\hat{\sigma}_{POL}(qq\to qq\gamma)}{d\hat{t}}=\frac{\hat{s}_1}{16\pi\hat{s}^4}|\bar{M}|_{POL}^2.$$

The differential cross-section at the hadron level is obtain expression used in (5)-(9).

İn the calculations of the differential cross section of processes with polarized protons, the polarized parton distribution functions PDF were used. The polarized parton distribution functions have the following form [26]:

$$\Delta u_v(x,Q_0^2)=x^{-0.441}(1-x)^{3.96}(0.928+0.149x^{0.5}-1.141x+11.612x^{1.5}),$$

$$\Delta d_v(x,Q_0^2)=x^{-0.665}(1-x)^{4.46}(-0.038-0.43x^{0.5}-5.260x+8.443x^{1.5}),$$

$$\Delta G(x,Q_0^2)=x^{-1.17}(1-x)^{5.33}(0.03-1.71x^{0.5}+3.01x+43.5x^{1.5}).$$

At the hadron level, the following relationship was obtained between the differential cross sections of collisions with longitudinally polarized and nonpolarized protons:

$$\frac{d\sigma_{pol}(qq\to qq\gamma)}{dt}=(1+\lambda_1\lambda_2)\frac{d\sigma(qq\to qq\gamma)}{dt}$$

The double spin asymmetry at the hadron level was calculated using formula

[25,27]:

$$A_{LL} = \frac{\sigma^{\uparrow\uparrow} - \sigma^{\uparrow\downarrow}}{\sigma^{\uparrow\uparrow} + \sigma^{\uparrow\downarrow}}, \quad (10)$$

where $\sigma^{\uparrow\uparrow}$ and $\sigma^{\uparrow\downarrow}$ – cross-section of processes, respectively, with the directed and opposite polarzed of colliding protons.

For the double spin asymmetry of the bremsstrahlung subprocess $qq \to qq\gamma$ at the parton level, the following formula was obtained:

$$\hat{A}_{LL} = \lambda_1 \lambda_2.$$

The differential cross sections of the bremsstrahlung subprocess $qq \to qq\gamma$ calculated in FeynCalc and simulated in PYTHIA 8.315 were compared.

PYTHIA was used with the following setup:

prompt photon production processes enabled:

PromptPhoton:all = on

minimum hard scale:

PhaseSpace:pTHatMin = 0.5 GeV

It is important to emphasize that in its default configuration, PYTHIA automatically includes:

- initial- and final-state parton showers (ISR/FSR),
- multiparton interactions (MPI),
- hadronization (Lund string model).

Therefore, the PYTHIA results contain contributions from non-perturbative effects, which are not included in our analytical calculations performed at the pure pQCD level (matrix elements for the $qq \to qq\gamma$ subprocess).

This explains the observed discrepancies in the low-$p_T$ region, where:

- soft radiation becomes important,
- hadronization and parton shower effects significantly modify the spectrum.

## III. NUMERICAL RESULTS AND THEIR DISCUSSION

### *3.1 Differential cross-section of the bremsstrahlung subprocess $qq \to qq\gamma$ of prompt photon production in collisions of nonpolarized protons*

Figure 1(a,b,c,d,e,f) shows the dependence of the total cross-section of prompt photons production in the subprocess of bremsstrahlung of a quark on a quark $qq \to qq\gamma$, at the hadron level on the sum of the energies of the colliding protons $\sqrt{s}$ and the dependence of the differential cross-section of the subprocess on the transverse momentum $p_T$, the cosine of the scattering angle $Cos(\theta)$, the rapidity of the photon y and $x_T$, calculated at collisions of nonpoloarized protons and schematically presented prompt photon production in bremsstrahlung.

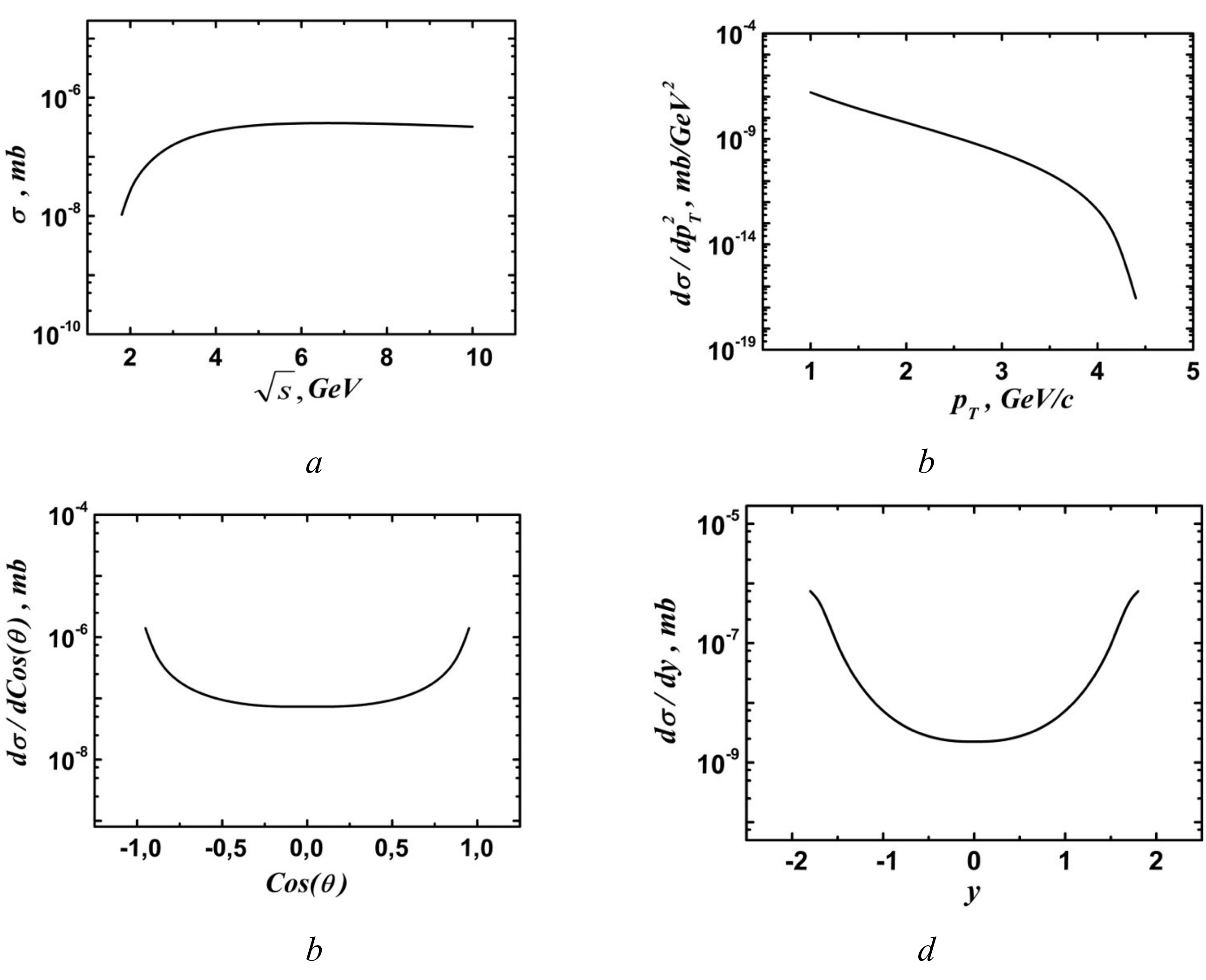

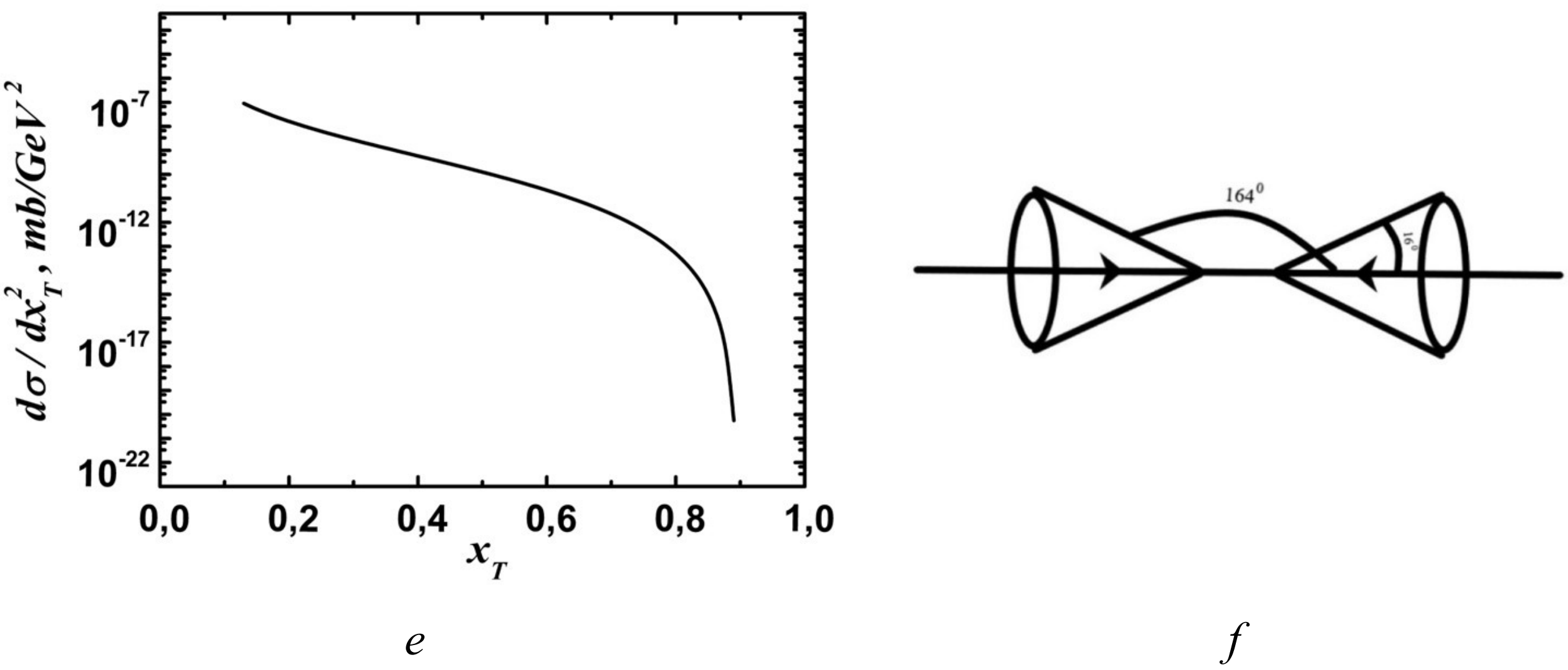

*e* *f*

Figure 1(a,b,c,d,e,f). The dependence of the total cross-section of prompt photons production in the bremsstrahlung of $qq \rightarrow qq\gamma$ on the sum of the energies of the colliding protons $\sqrt{s}$ (a) and the dependence of the differential cross-section of the subprocess on the transverse momentum $p_T$ (b), the cosine of the scattering angle $Cos(\theta)$ (c), the rapidity of the photon y (d) and $x_T$ (e ), at nonpolarized protons collisions and schematically presentation of prompt photon production in bremsstrahlung (f)

As can be seen from Figure 1(a), at low energies ($\sqrt{s}$<3 GeV), the differential cross-section is very small and increases rapidly with increasing $\sqrt{s}$. This is because photon production is suppressed by the strong coupling constant $\alpha_s$. Photon production is dominated by soft bremsstrahlung, in which the photon carries a small fraction of the quark energy. Moreover, it is known that, the result of bremsstrahlung is linearly proportional to the square of the particle acceleration. At high proton collisions energies, the partons in the proton have a high velocity, therefore, the sharp decrease in velocity due to braking will be significant.

At intermediate energies (3<$\sqrt{s}$<7 GeV), the differential cross-section reaches a maximum and then slowly decreases with increasing $\sqrt{s}$. Photon production is enhanced by collinear bremsstrahlung, in which the photon is produced in the direction of the quark. Collinear enhancement occurs in dense media due to interference between different scattering sites, also known as the Landau-Pomeranchuk-Migdal effect [28-31]. This interference can increase the particle's wavelength if the longitudinal momentum transfer becomes small. If the wavelength becomes larger than the mean free path in the medium (the average distance between scattering sites), the scatterings can no longer be considered as independent events. This results in a suppressed production spectrum compared to that predicted by the Bethe-Heitler formula, which assumes independent scattering [28-31]. In QED, low photon energies are suppressed due to the LPM effect, while in QCD, high gluon energies are suppressed. The effect was experimentally confirmed in 1994 at the Stanford Linear Accelerator Center (SLAC, California, USA). Photon emission is also affected by the change in the coupling constant $g_s$, which decreases with increasing energy.

In the high-energy region ($\sqrt{s}$>7 GeV), the differential cross-section decreases rapidly with increasing $\sqrt{s}$. This is because photon production is suppressed, the photon carrying most of the quark energy. The hard suppression is due to the "dead" cone effect [28-31] in which photon production is forbidden within a small angle around the quark direction. The "dead" cone effect is caused by the quark mass, which reduces the phase space for photon producion. Figure 1(f) shows as photons are maximally produced in the direction of the cone surface. Photons are produced at angles close to the particle collision axis (16 and 164 degrees).

The dependence shown in Figure 2(b) is a logarithmic plot of the differential cross-section versus the photon transverse momentum $p_T$, which shows that the differential cross-section decreases rapidly as the transverse momentum $p_T$ increases. This is because the probability of production a high-energy photon in a quark-quark collisions is very small, since the photon carries away most of the energy and momentum of the original quarks. The plot also shows that the differential cross-section tends to zero as $p_T$ approaches the kinematic limit $p_T = \sqrt{s}/2$. This is because photon production becomes kinematically forbidden when the final quarks have zero energy and momentum. The graph can be compared with the theoretical predictions of the Broten-Pisarski method in thermal QCD, which calculates the rate of photon production from the QGP in equilibrium of the medium [32,33]. The method includes coherence effects between different scattering sites, also known as the LPM effect, which reduces the photon production rate at high energies. The method also explains the nonequilibrium nature of the QGP, which can be characterized by three constants: reflecting the degree of isotropy and viscosity of the medium [31,32].

From the dependence of differential cross-section on cosine scattering angle of photon $Cos(\theta)$ (Figure 1(c)) it is evident that the differential cross-section has a symmetrical bell-shaped curve relative to the cosine of the scattering angle of promptt photons. This means that the probability of photon production is the lowest when quarks fly apart at a right angle ($Cos(\theta) = 0$), and the highest when they scatter in the same or opposite direction ($Cos(\theta) = \pm 1$). A similar picture is observed in the dependence of the differential cross-section on rapidity y prompt photon (Figure 1(d)).

The dependence of the differential cross-section on the parameter $x_T$ (Figure 1(e)) shows that the differential cross-section decreases rapidly with increasing $x_T$. This means that the probability of production a photon with a high transverse momentum $p_T$is very small compared to a photon with a low transverse momentum $p_T$. Which is consistent with the dependence of the differential cross-section on the transverse momentum of the produced prompt photons $p_T$. This is due to the fact that photons with a high transverse momentum $p_T$ are suppressed by the LPM effect, which is a quantum interference phenomenon that reduces the radiation of a charged particle crossing a medium. The LPM effect is more pronounced for quark bremsstrahlung than for quark-antiquark annihilation, which is another subprocess of photon production in the QGP. Therefore, bremsstrahlung photons are predominantly soft, i.e. they have a small transverse momentum $p_T$.

The analysis showed that, in bremsstrahlung at low energies proton collisions are more likely to produce photons with a large transverse momentum $p_T$, and at high energies - collisions of photons with a small transverse momentum $p_T$. The comparision showed, in the Compton scattering quark-gluon at low energies proton collisions the production of photons with a small $p_T$ is more likely, and at high energies proton collisions the production of photons with a large $p_T$ is less likely.

*3.2 Differential cross-section of the bremsstrahlung subprocess $qq \to qq\gamma$ of prompt photon production in collisions of longitudinally polarized protons*

Figure 2(a,b,c,d,e) shows the dependence of the total cross-section of the subprocess $qq \to qq\gamma$ taking into account the longitudinal polarization of the colliding protons at polarization order $P_1$, $P_2$=0.9, -0.9 (in that $P_1P_2$=-0.81) and at $P_1$=$P_2$=±0.9 (in that $P_1P_2$==0.81) on the sum of the energy of the colliding particles $\sqrt{s}$ and the dependence of the differential cross-section of the subprocess on the transverse momentum of promptt photons $p_T$, the cosine of the scattering angle, the rapidity of prompt photons y and the parameter $x_T$ at $\sqrt{s}$=10 GeV.

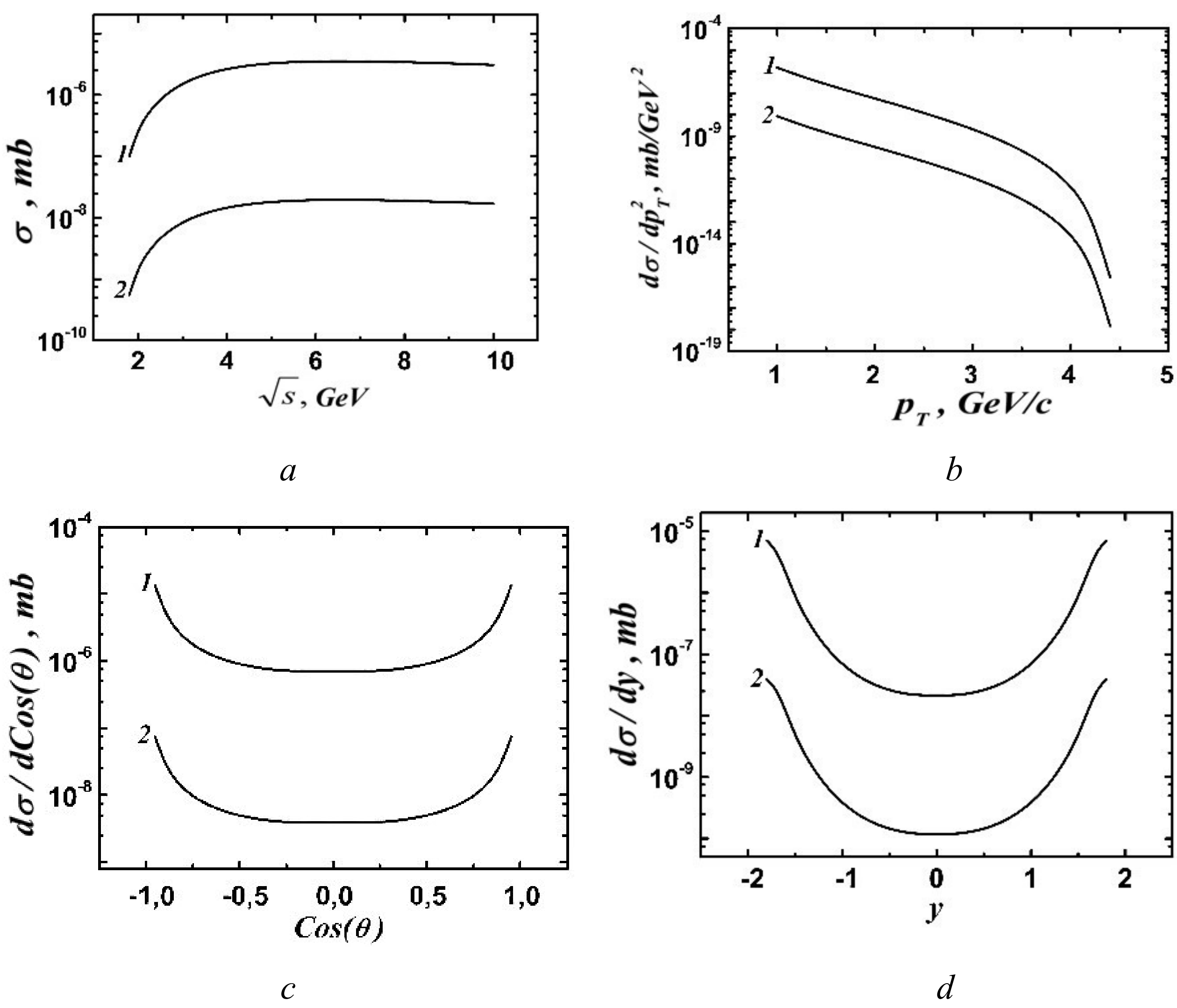

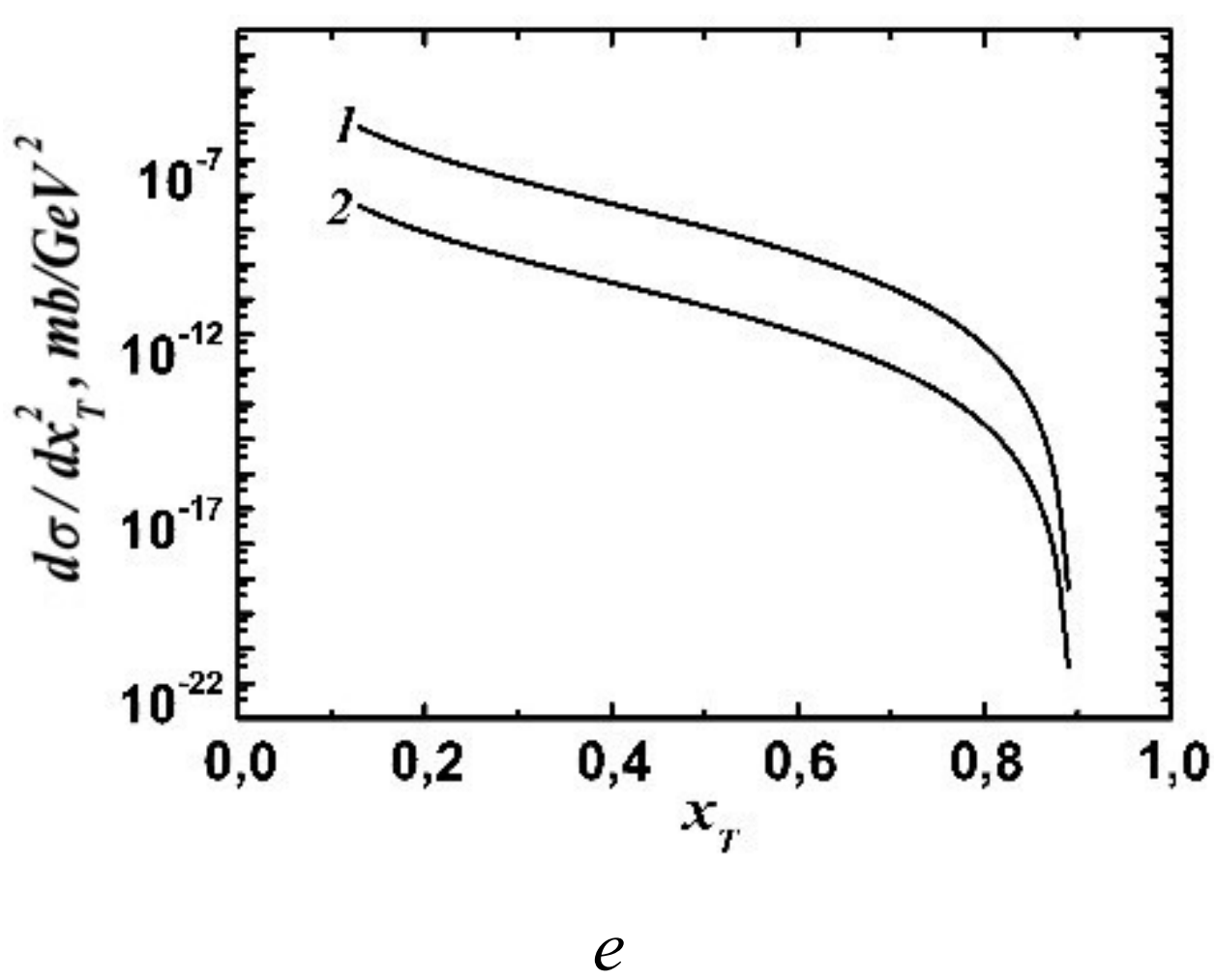

*e*

Figure 2(a,b,c,d,e). Dependence of the total cross-section of the subprocess $qq \to qq\gamma$ taking into account the polarization of colliding protons at P$_1$, P$_2$=0.9, -0.9 (in that $P_1P_2$=-0.81) (curve 2) and at $P_1$=$P_2$=±0.9 (in that $P_1P_2$=0.81) (curve 1) on the sum of the energy of the colliding particles $\sqrt{s}$ (a) and the dependence of the differential cross-section of the subprocess on the transverse momentum of prompt photons $p_T$ (b), the cosine of the scattering angle (c), the rapidity of prompt photons (d) and the parameter $x_T$ (e) at $\sqrt{s}$=10 GeV

As can be seen from the dependence of the total cross-section taking into account the longitudinal polarization of protons on the sum of the energy of the colliding particles $\sqrt{s}$ (Figure 2(a)), the influence of polarization is the practically same over the entire range of change in the energy of the colliding particles.

As can be seen from Figure 2(b), the dependence of the differential cross-section on the photon transverse momentum $p_T$ is due to polarization having a large effect at small values of the photon transverse momentum $p_T$. As the photon transverse momentum $p_T$ increases, the effect of polarization on the differential cross-section decreases. For bremsstrahlung $qq \to qq\gamma$, the effect of polarization can lead to either an increase or a decrease in the differential cross-section of this subprocess, depending on the spin orientation of the initial particles. When the spin orientations are different, the total cross-section decreases (curve 2), while with the same orientation it increases (curve 1), compared to the nonpolarized case. The polarization effect can be qualitatively understood by considering the interference between the colliding particles, which depends on the mutual orientation of their spins and determines the phase difference of the radiation waves. When the spins are parallel, the phase difference is zero and the interference is constructive. When the spins are antiparallel, the phase difference is $\pi$ and the interference is destructive. The polarization effect can also be quantified by calculating the polarization tensor, which is a function describing the modification of the photon propagator due to the presence of the medium. The polarization tensor depends on the frequency, wave vector, and magnetic field of the photon, as well as on the temperature, chemical potential, and polarization of the medium. The polarization tensor determines the dispersion law, the attenuation rate, and the polarization state of the photon. The differential cross-section of bremsstrahlung is proportional to the imaginary part of the polarization tensor. Therefore, the polarization effect can lead to either an increase or a decrease in the differential cross-section, depending on the system parameters.

The Figure 2(c) shows the predicted angular dependence for one of the important, though usually non-dominant subprocesses. Comparison with data (after accounting for all other contributions) can help isolate the contribution from this speciic channel.

The Figure 2(d) shows as these data are essential for convoluting the partonic cross-sections with the parton distribution functions when calculating the total cross-section for hadronic collisions.

Figure 2(e) provides an important theoretical input for the complete analysis of spin asymmetries in $pp$ collisions, where the $qq \to qq\gamma$ subprocess can give a noticeable contribution, especially in certain kinematic regions. Comparing with data can help determine the relative importance of this process compared to the dominant Compton scattering process, $qg \to q\gamma$.

We has been investigated the dependence double spin asymmetry $A_{LL}$ on kinamtic parameters: transverse mpmentum $p_T$ of photon and $x_T$. In the Figure 3(a,b) are presented the dependence of double spin assymmetery $A_{LL}$ on transverse momentum $p_T$ of photon and $x_T$.

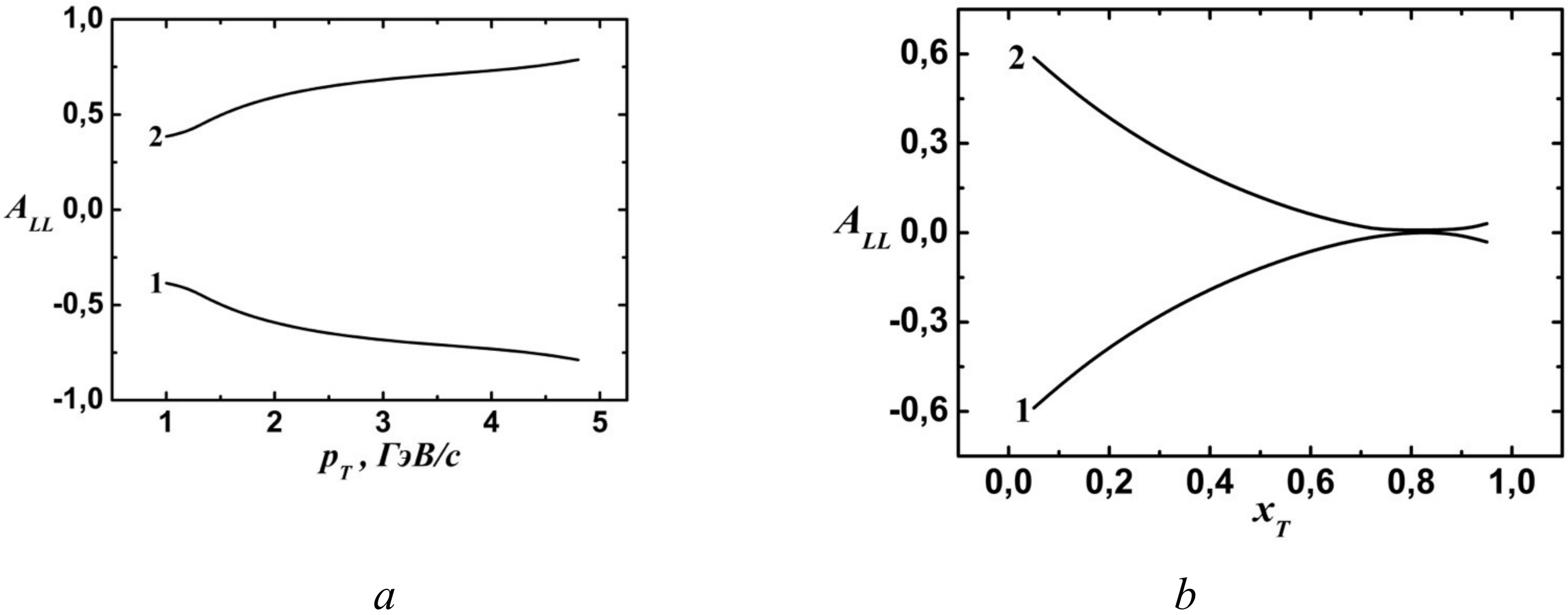

*a* *b*

Figure 3(a,b). The dependence of double spin assymmetery $A_{LL}$ on transverse momentum $p_T$ photon (a) and $x_T$ (b)

The sign of the asymmetry indicates which proton spin configuration is preferred for this process (Figure 3(a)). Curve 1, $A_{LL}$<0 means that for this configuration, the cross-section with antiparallel spins is larger than with parallel spins ($\sigma^{\uparrow\downarrow}>\sigma^{\uparrow\uparrow}$), Curve 2, $A_{LL}$>0 - means the opposite ($\sigma^{\uparrow\uparrow}>\sigma^{\uparrow\downarrow}$). The most important observation here is that the magnitude of the asymmetry $|A_{LL}|$ increases with increasing $p_T$. This means that the spin dependence of the $qq \rightarrow qq\gamma$ subprocess becomes stronger for "harder" collisions (those with larger transverse momentum transfer). As $p_T$ grows, the interaction becomes more sensitive to the relative spin orientation of the initial protons.

This is the most significant feature of the graph (Figure 3(b)). The existence of such a point where the asymmetry vanishes is a non-trivial prediction of pQCD. It means that at a specific ratio of kinematic variables (specifically, at $x_T \approx$ 0.7-0.75), the $qq \rightarrow qq\gamma$ process becomes completely insensitive to the spin configuration of the initial quarks. The cross-sections for parallel and antiparallel spins become equal. This "zero-asymmetry" effect arises from the interference and mutual cancellation of different Feynman diagrams that describe this subprocess. The contributions of these diagrams to the spin-dependent part of the cross-section have different dependencies on kinematic variables, and at one specific point, their contributions can exactly cancel each other out.

Figure 4(a,b) shows the dependence of the double-spin asymmetry $A_{LL}$ of the $qq \to qq\gamma$ subprocess on the product of the polarization orders $P_1$ and $P_2$ at $\sqrt{s}$=10 GeV and 3D graphic of dependence of double-spin asymmetry $A_{LL}$ on polarization orders $P_1$ and $P_2$.

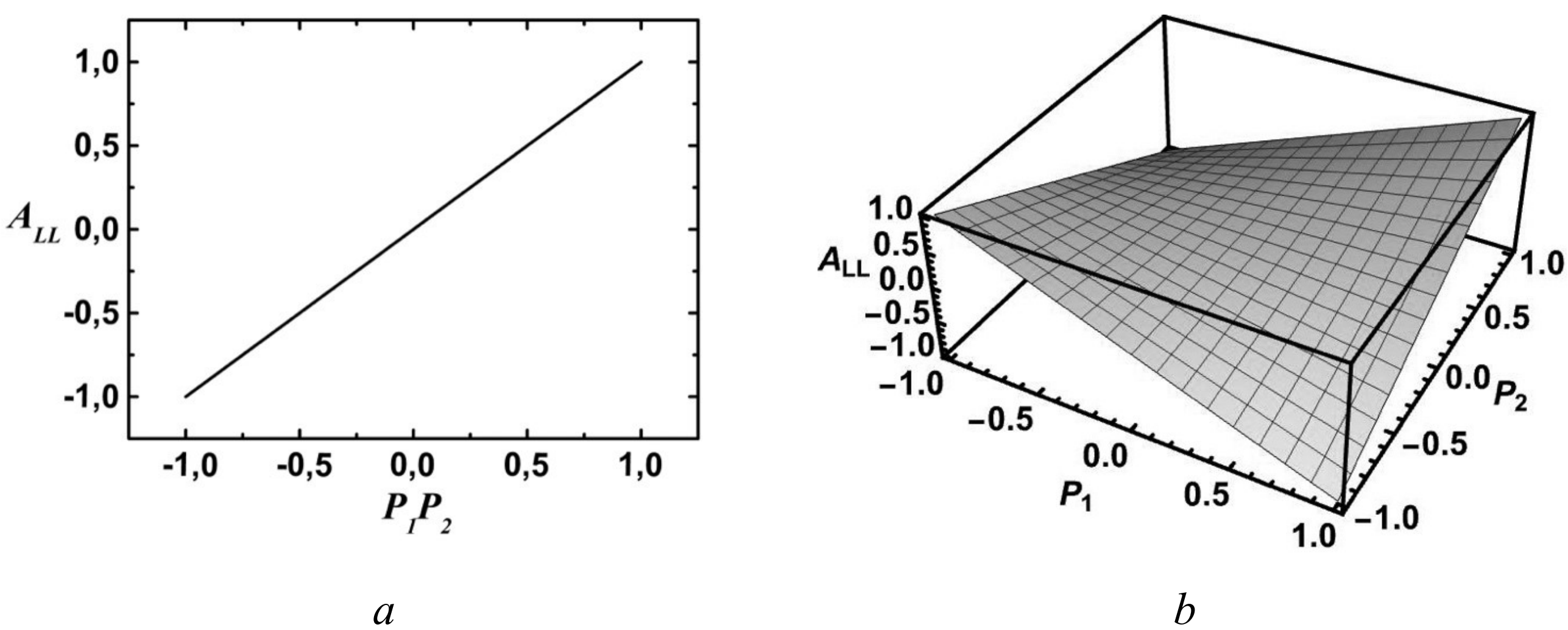

*a* *b*

Figure 4(a,b). Dependence of the double-spin asymmetry $A_{LL}$ of the $qq \to qq\gamma$ subprocess on the product of the polarization orders $P_1P_2$ at $\sqrt{s}$=10 GeV (a), 3D graphic of dependence of double-spin asymmetry $A_{LL}$ on polarization orders $P_1$ and $P_2$ (b)

Full polarization of both protons beams (e.g., $P_1 = P_2 = \pm 1$, that is $P_1P_2 = 1$) leads to maximal double spin asymmetries $A_{LL} = 1$. At opposite values of polarization of the colliding quarks ($P_1 = 1, P_2 = -1$ or $P_1 = -1, P_2 = 1$, that is $P_1P_2 = -1$), the minimum value of the double spin asymmetry $A_{LL} = -1$ is achieved. No polarization in at least one beam ($P_1 = 0$ or $P_2 = 0$) results in zero double spin asymmetry $A_{LL} = 0$ (see Figure 4(a,b)). This behavior reflects a strong spin-dependence of the underlying partonic process, likely due to interference terms in the matrix elements sensitive to spin orientations.

Such a linear relation is crucial for spin experiments, peculiarity at NICA SPD, as it enables direct interpretation of measured $A_{LL}$ values in terms of beam polarizations and the underlying QCD dynamics.

*3.3 Comparison of differential cross-sections of the subprocesses bremsstrahlung $qq \to qq\gamma$*

Figure 5(a,b,c) shows the dependences of the ratio $R = \frac{d\sigma_{POL}(qq\to qq\gamma)}{d\sigma(qq\to qq\gamma)}$ at the hadron level taking into account of the polarization order of the initial particles with $P_1$, $P_2$=0.9, -0.9 on the sum of the energy of the colliding particles $\sqrt{s}$, the transverse momentum of prompt photons $p_T$, and $x_T$ at $\sqrt{s}$=10 GeV.

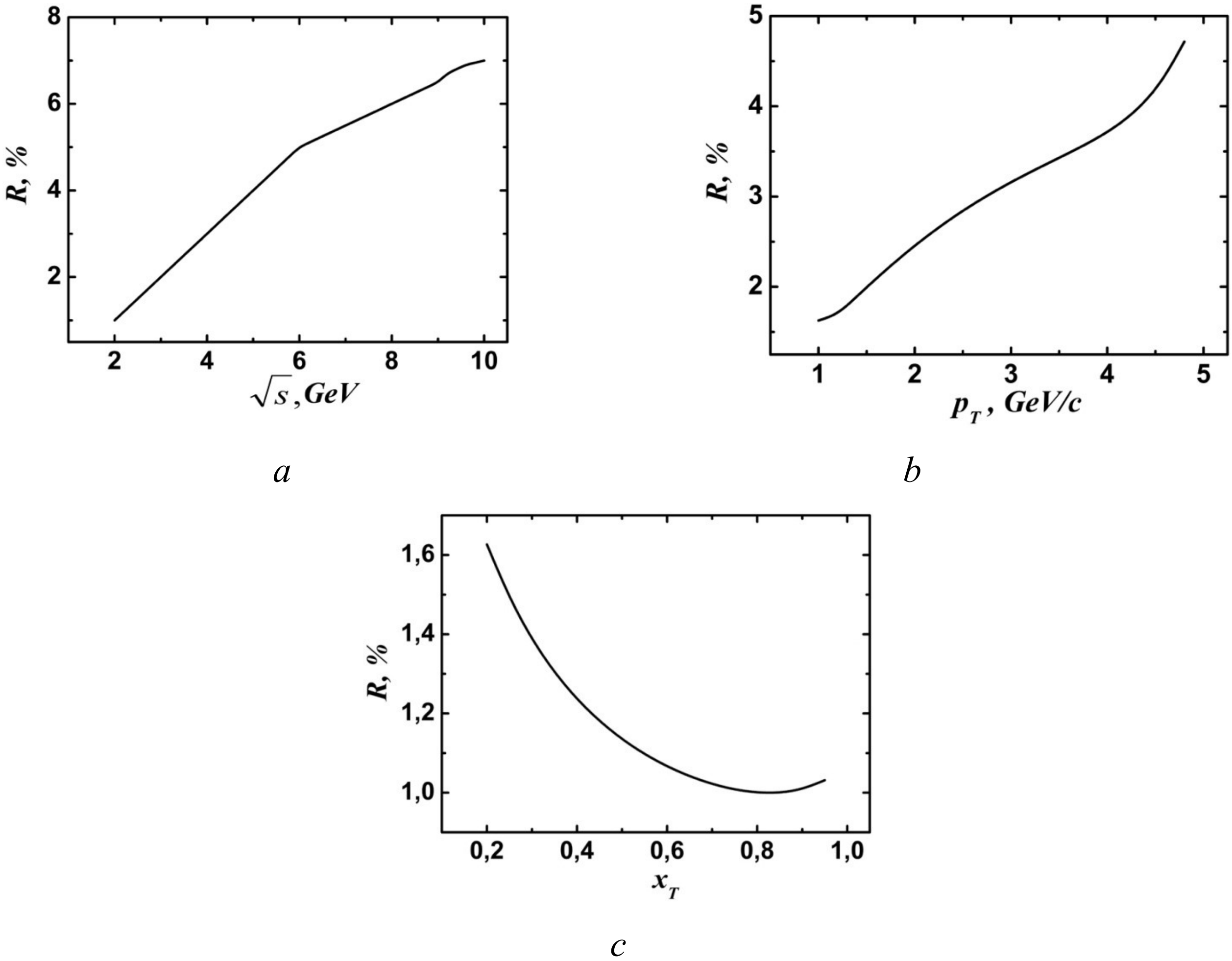

Figure 5(a,b,c). Dependence of the ratio $R = \frac{d\sigma_{POL}(qq\to qq\gamma)}{d\sigma(qq\to qq\gamma)}$ at the hadron level taking into account of the polarization order of the initial particles with $P_1, P_2$=0.9, -0.9 on the sum of the energy of the colliding particles $\sqrt{s}$ (a), the transverse momentum of prompt photons $p_T$ (b) and $x_T$ (c) at $\sqrt{s}$=10 GeV

As seen from Figure 5(a) with increasing collision energy $\sqrt{s}$, the value of $R$

also increases. This indicates that polarization effects become more significant at higher energies. This happens because quantum spin effects and spin correlations become more pronounced at high energies, especially in processes where photon production is involved.

As seen from Figure 5(b) similarly, as the transverse momentum $p_T$ of the photon increases, so does $R$. Photons with high $p_T$ typically originate from hard processes that are more sensitive to the spin states of the initial particles. Thus, polarization effects play a more prominent role in these events.

Both plots (Figure 5(a,b)) show that polarization effects are enhanced at higher energies and for photons with greater transverse momenta $p_T$.

Figure 5(c) show, as $R$ is a measure of the relative importance or magnitude of spin effects in the $qq \to qq\gamma$ subprocess. A value of $R = 1.5\%$ means that the spin-dependent part of the cross-section is 1.5% of the magnitude of the spin-independent (averaged) part. The calculation is performed for a hadron collisions (likely $pp$ or $p\bar{p}$) with antiparallel polarization of the initial particles ($P_1$=0.9, $P_2$=-0.9) at an energy of $\sqrt{s}$=10 GeV. The graph shows how the relative contribution of spin effects from the $qq \to qq\gamma$ subprocess changes depending on the hardness of the collision.

The decrease in $R$ (from $x_T$≈0.2 to 0.8) means that as $x_T$ increases, the spin-dependent part of the cross-section ($d\sigma_{POL}$) decreases faster than the spin-independent part ($d\sigma$), or it grows more slowly (although both cross-sections are likely falling). As a result, their ratio decreases. This suggests that in this kinematic region, spin effects become relatively less prominent compared to the total cross-section.

Minimum at $x_T$≈0.8, this point represents a region where the relative contribution of spin effects from this subprocess is at its minimum. This behavior is a non-trivial result arising from the complex convolution of PDFs (both polarized $q_{POL}(x)$ and nonpolarized $q(x)$) with the partonic cross-sections. It's likely that at large $x$ (which corresponds to large $x_T$), the ratio of parton polarization $q_{POL}(x)/q(x)$ decreases, leading to the reduction in $R$.

The rise ($x_T$>0.8) at the very edge of the phase space (when the photon carries away almost all the available energy) is a subtle effect. It might indicate that in this extreme kinematic regime, the spin-dependent parts of the cross-section start to fall more slowly than the spin-independent parts, thus increasing their ratio again. This region is very sensitive to the behavior of parton distributions as $x \to 1$.

*3.4 Comparison of the differential cross sections of the qq→qqγ bremsstrahlung subprocesses obtained by PYTHIA simulations and calculated in FeynCalc*

The dependences of the differential cross section on the sum of the colliding protons $\sqrt{s}$, the photon transverse momentum $p_T$, the photon scattering angle cosine $Cos(\theta)$, and the rapidity $y$ were obtained by simulating the process in PYTHIA 8.315 [34].

Figure 6(a,b,c,d) shows the dependences of the differential cross section on the sum of the colliding proton energies $\sqrt{s}$, the transverse momentum of the prompt photon $p_T$, the photon scattering angle cosine $Cos(\theta)$, and the rapidity $y$.

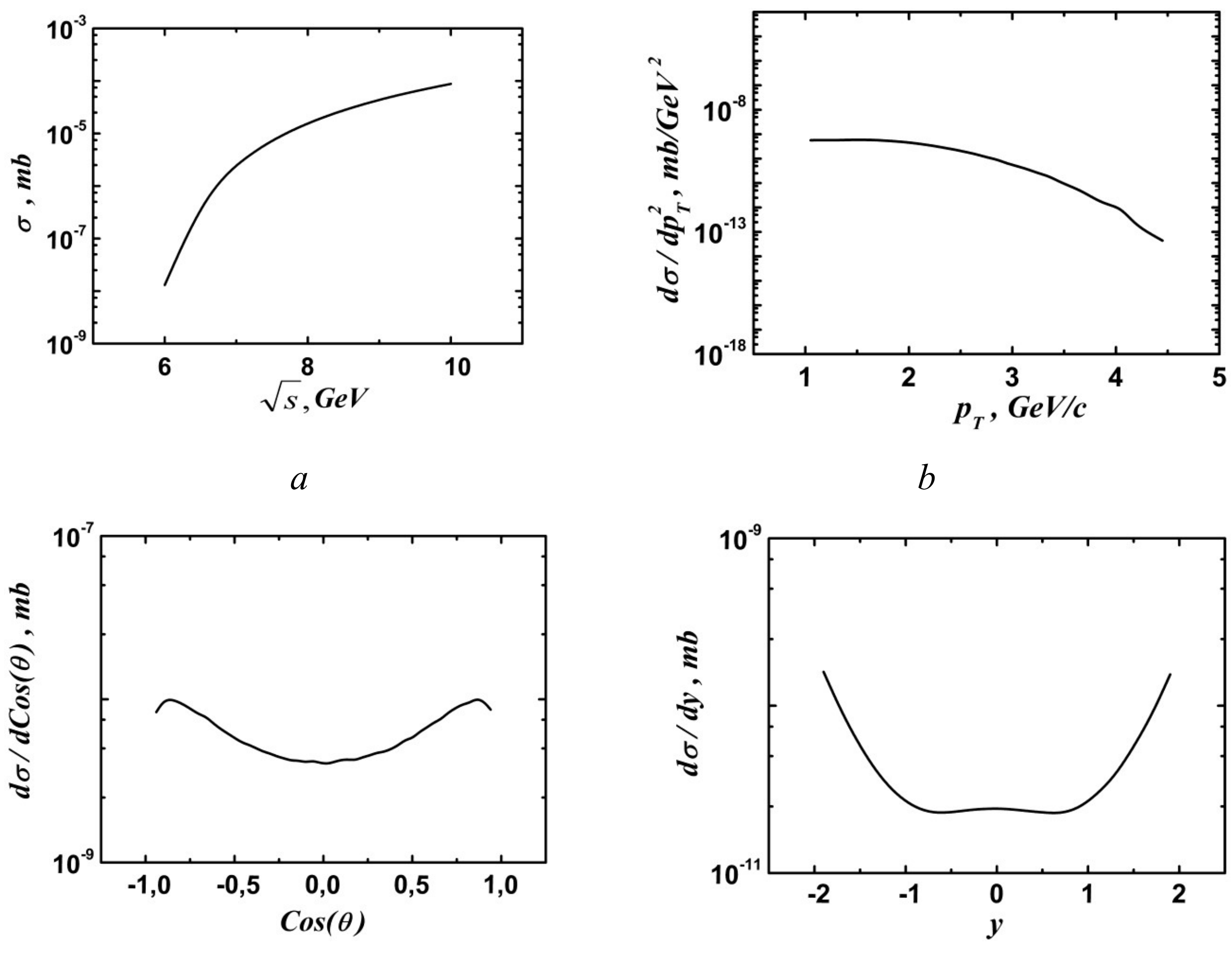

*c* *d*

Figure 6(a,b,c,d). Dependences of the differential cross section on the sum of the colliding protons energies $\sqrt{s}$ (a), the transverse momentum of the prompt photon $p_T$ (b), the cosine of the photon scattering angle $Cos(\theta)$ (c), and the rapidity $y$ (d).

As can be seen from Figure 6(a,b,c,d), the dependences of the differential cross section on the sum energy of colliding protons $\sqrt{s}$, the transverse momentum of the prompt photon $p_T$, the cosine of the photon scattering angle $Cos(\theta)$, and the rapidity $y$, obtained using PYTHIA 8.315 and calculated in FeynCalc, are identical. They differ in their numerical values.

The dependences of the ratio $R = \frac{\sigma(qq \to qq\gamma)_{FeynCalc}}{\sigma(qq \to qq\gamma)_{PYTHIA}}$ on the sum energy of colliding protons $\sqrt{s}$, the transverse momentum of the photon $p_T$, the cosine of the photon scattering angle $Cos(\theta)$ and the rapidity $y$ are shown in Figure 7(a,b,c,d).

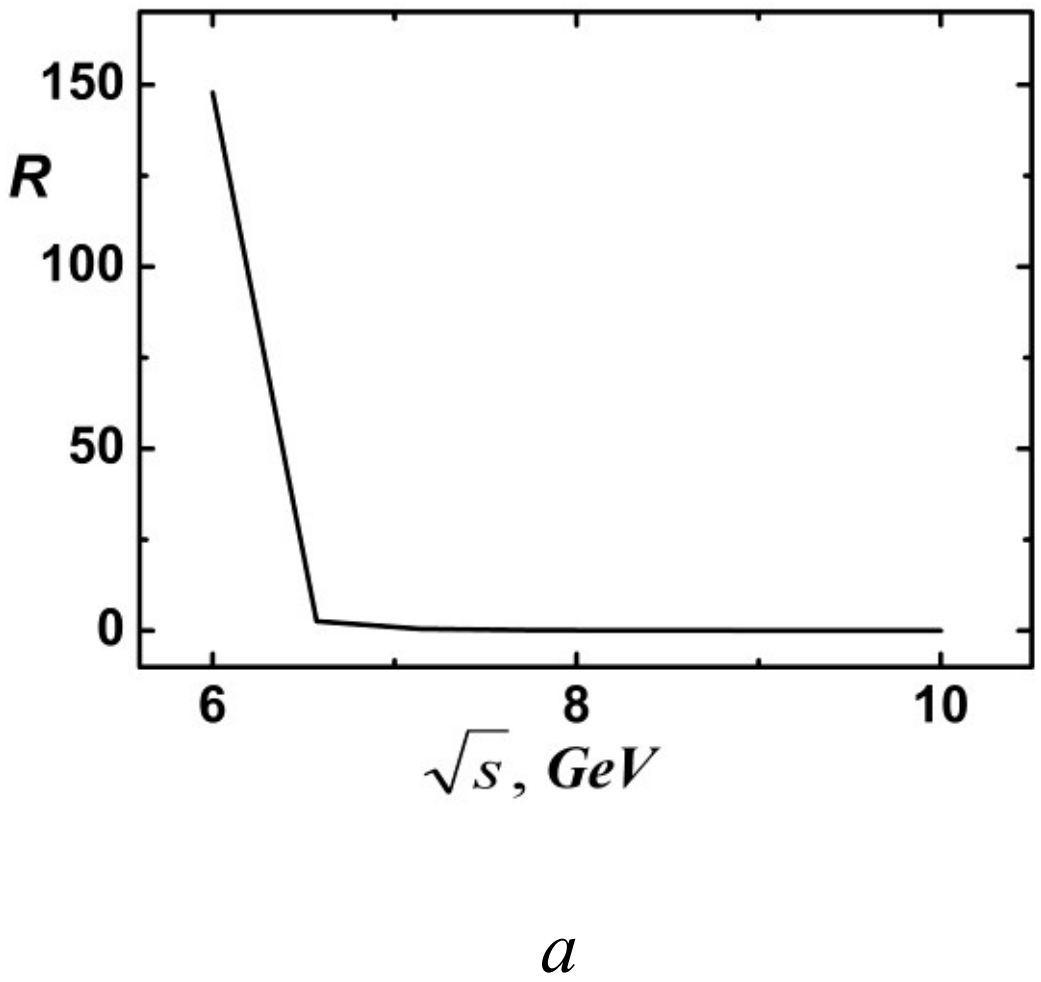

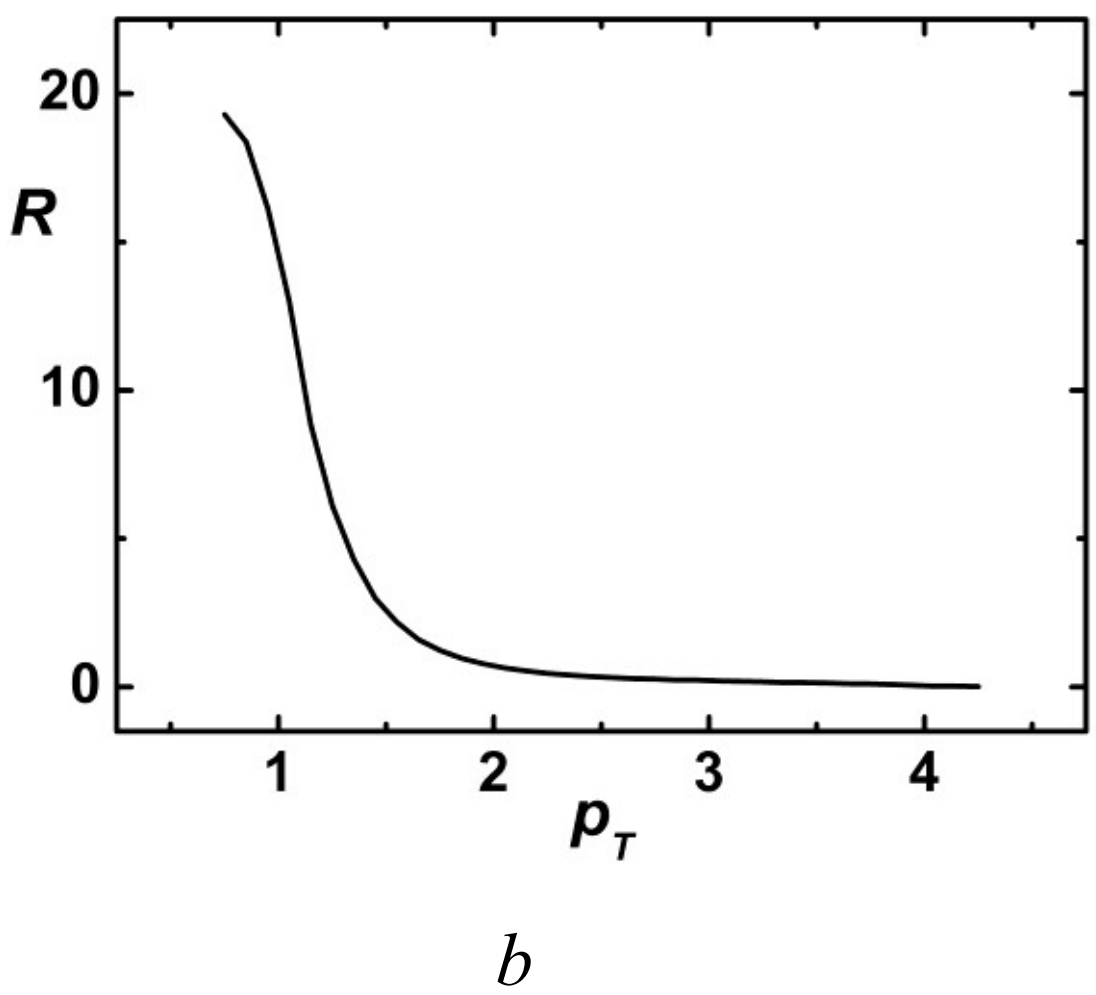

*a* *b*

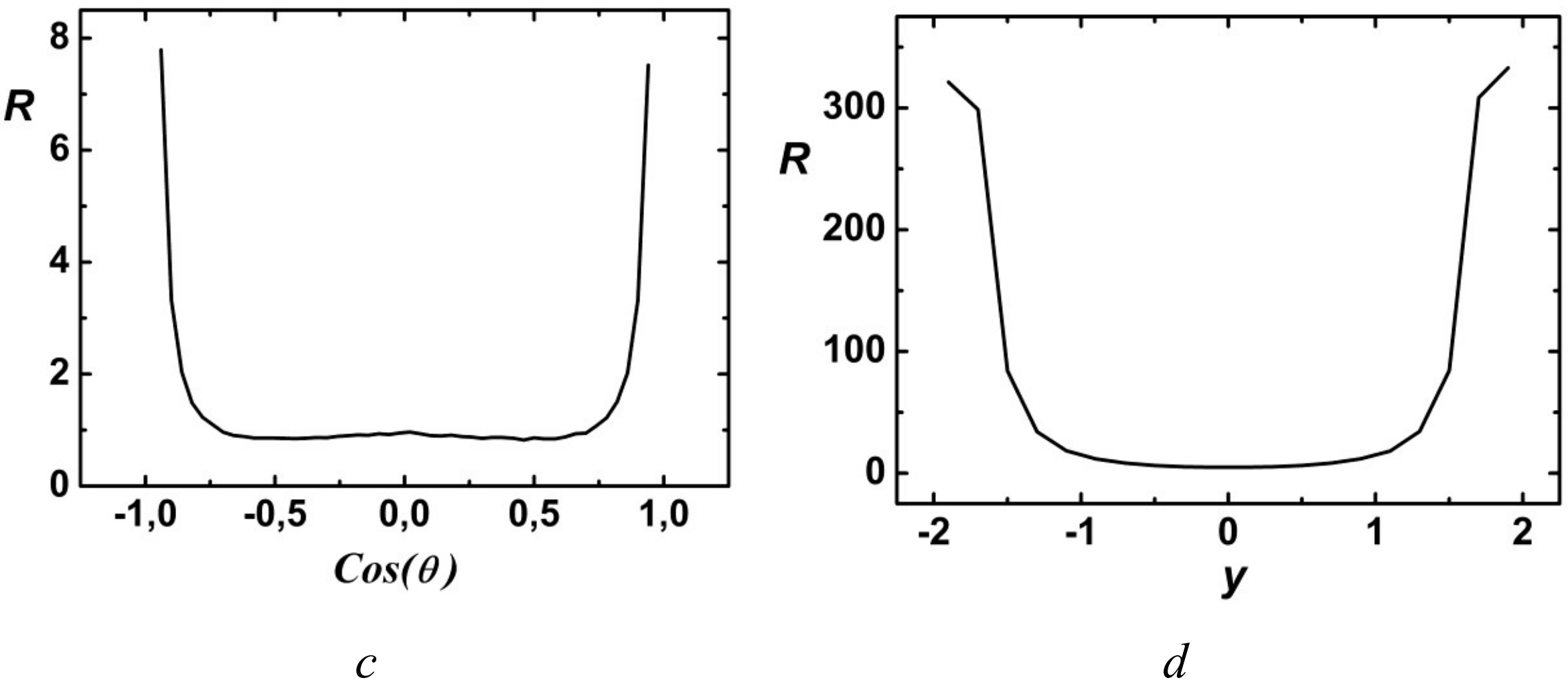

*c* *d*

Figure 7(a,b,c,d) Dependences of the ratio $R = \frac{\sigma(qq \to qq\gamma)_{FeynCalc}}{\sigma(qq \to qq\gamma)_{PYTHIA}}$ on the sum energy of colliding protons $\sqrt{s}$ (a), photon transverse momentum $p_T$ (b), photon scattering angle cosine $Cos(\theta)$ (c), and rapidity $y$ (d).

As can be seen from Figure 7(a,b), the differential dependences of the cross section on the sum energy of colliding protons $\sqrt{s}$ and photon transverse momentum $p_T$, obtained in PYTHIA simulations at lower energies $\sqrt{s}$ and photon transverse momentum $p_T$, are lower than similar dependences calculated in FeynCalc. In contrast, the PYTHIA 8 results include: initial and final parton cascades (ISR/FSR), soft and collinear radiation, multiple parton interactions (MPI), and hadronization (Lund string model).

This difference becomes particularly significant at low energies ($\sqrt{s} \approx 10$ GeV), which follows directly from the physics of the process:

1. Enhancement of non-perturbative effects

At low energies, the contribution of non-perturbative QCD increases because: the characteristic scales of the process become comparable to $\Lambda_{QCD}$ (the fundamental scale of quantum chromodynamics, which defines the boundary between: the perturbative regime (high energies, weak coupling),the nonperturbative regime (low energies, strong coupling)), and the effects of hadronization and soft radiation are enhanced.

2. Dominance of soft kinematics

In the region of small $p_T$: bremsstrahlung has an infrared-sensitive character, and parton cascades in PYTHIA significantly modify the photon spectrum.

3. Limitations of the pure parton approximation

As shown in the article, the contribution of the qq→qqγ process accounts for only ~0.03% of the total cross section for prompt photon production. Consequently, the analytical calculation describes a narrow component of the process, whereas PYTHIA simulates the entire set of photon production mechanisms.

## CONCLUSIONS

Carried out investigation, showed, that, at NICA energies, the differential cross-section of the prompt photon production process at proton-proton collisions consists of more than 50% of the differential cross-section of Compton scattering of quark-gluon $qg \to q\gamma$ subprocess, about 43% of the differential cross-section of quark-antiquark pair $q\bar{q} \to g\gamma$ annihilation and less than 0.03% of the differential cross-section of bremsstrahlung $qq \to qq\gamma$ .

Comparision of obtained results, showed, that in bremsstrahlung at low energies proton collisions are more likely to produce photons with a large transverse momentum $p_T$, and at high energies - collisions of photons with a small transverse momentum $p_T$. In the Compton scattering quark-gluon at low energies the production of photons with a small $p_T$ is more probable in the collisions of protons, and at high energies the production of photons with a large $p_T$ is less probable.

Photons are produced at angles close to the particle collision axis (16 and 164 degrees).

Bremsstrahlung is an electromagnetic process in which a photon combines with the electric charge of a particle. Annihilation of a quark-antiquark pair is a strong process associated with color charge. The effect of polarization in bremsstrahlung is due to interference (LMP effect), and in annihilation - to conservation of angular momentum.

Taking into account longitudinal polarization does not change the nature of the dependence of differential cross sections on kinematic parameters. It can increase or decrease the values of the differential cross section. If the polarization of colliding particles is directed in one direction, the differential cross section decreases, in the opposite direction, the differential cross section increases. Polarization has a stronger effect at high collisions energies, since relativistic protons become more sensitive to the photon and gluon fields. Its effect is maximum at small $p_T$ , and decreases with increasing $p_T$.

Considering the previous results, we can conclude that the influence of polarization on the subprocess of quark-antiquark annihilation $q\bar{q} \rightarrow g\gamma$ is greater than on the subprocesses of Compton scattering $qg \rightarrow q\gamma$ and bremsstrahlung $qq \rightarrow qq\gamma$.

Polarization is significant at small $p_T$ for Compton scattering of a quark-gluon $qg \rightarrow q\gamma$ and annihilation of a quark-antiquark pair $q\bar{q} \rightarrow g\gamma$. For bremsstrahlung $qq \rightarrow qq\gamma$ its influence of polarization is noticeable at large $p_T$.

The dependence of the double-spin asymmetry on the polarization of colliding particles for the bremsstrahlung is $\hat{A}_{LL} = \lambda_1\lambda_2$, and quark-antiquark annihilation - $\hat{A}_{LL} = -\lambda_1\lambda_2$ subprocesses differs in sign.

Character of the dependences of differential cross-section on sum energy of colliding protons $\sqrt{s}$, transverse momentum of prompt photon $p_T$, cosine of scattering angle photon *Cos(θ)* and rapidity $y$ obtained by PYTHIA 8.315 and calculated in FeynCalc are identical. They differ in numerical values. The differential cross-section dependencies on sum energy colliding protons $\sqrt{s}$ and transverse momentum of photon $p_T$ obtained during modeling in PYTHIA 8.315 at lower energies $\sqrt{s}$ and transverse momentum of photon $p_T$ are lower than the analogous dependencies calculated in FeynCalc. PYTHIA provides a more realistic description of low-energy processes by incorporating additional non-perturbative effects, such as parton cascades, multiple interactions, and hadronization. For higher energy $\sqrt{s}$ and transverse momentum $p_T$ values, the differential cross-section dependences found in the PYTHIA simulation and calculated in FeynCalc coincide.

## Appendix

In the Figure(a,b,c,d,e,f,g,h,i,j,k,l,m,n,o,p) are shown Feynman diagrams of the prompt photons production in bremsstrahlung $qq \to qq\gamma$.

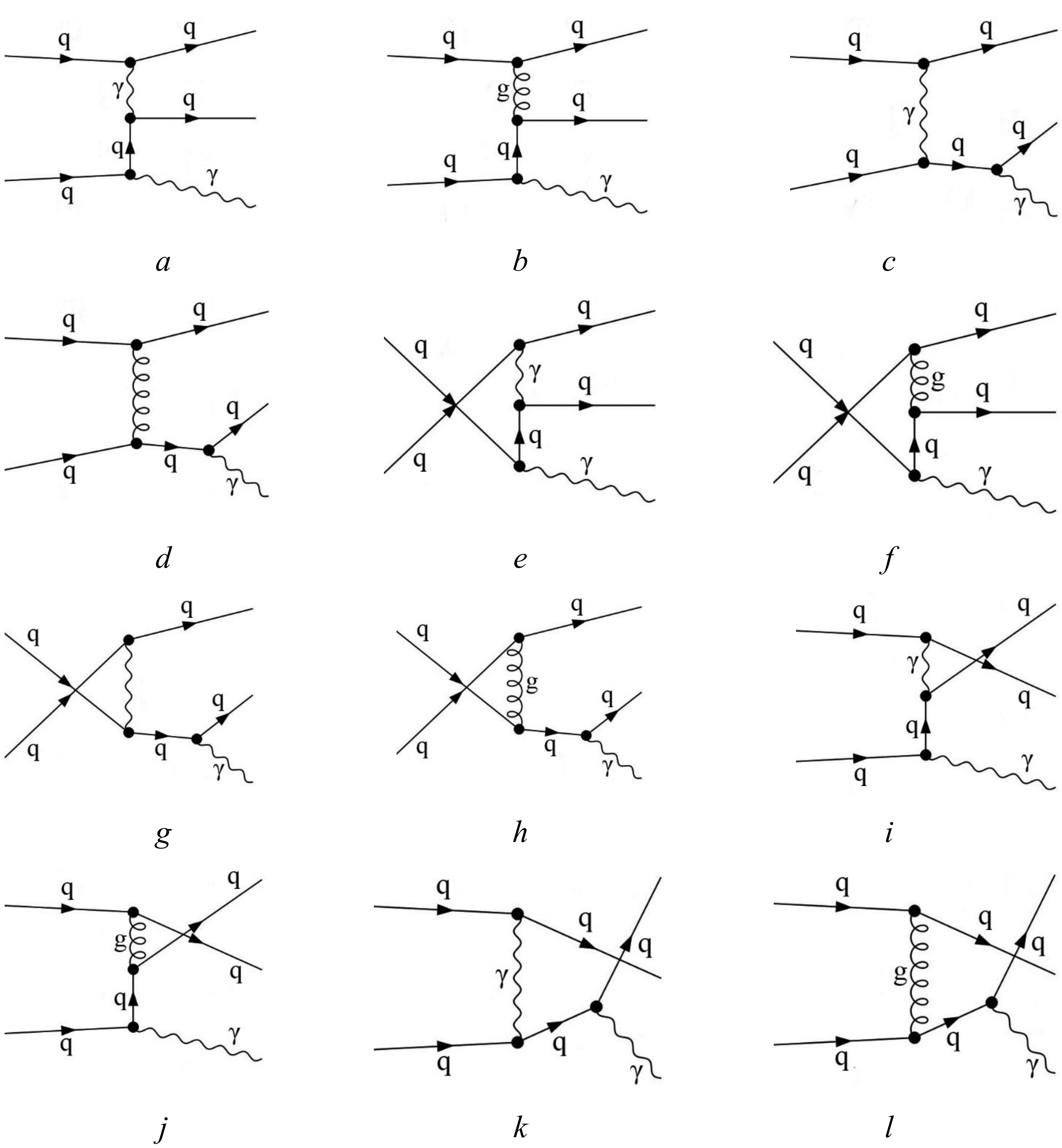

*a* *b* *c* *d* *e* *f* *g* *h* *i* *j* *k* *l*

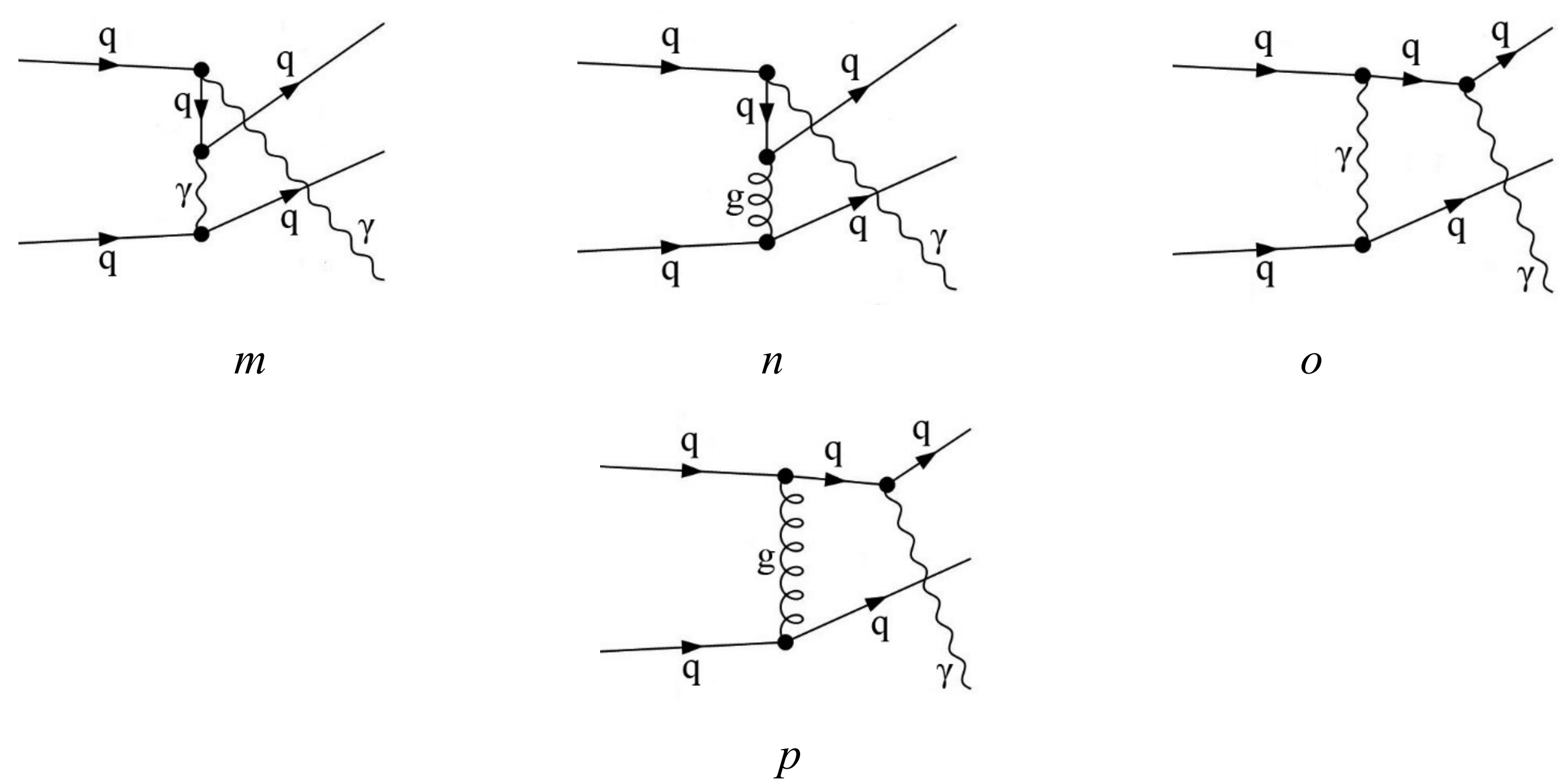

Figure(a,b,c,d,e,f,g,h,i,j,k,l,m,n,o,p). Feynman diagrams of the bremsstrahlung subprocess $qq \to qq\gamma$ of prompt photon production in collisions protons

The Mandelstam invariants for the subprocess $q(p_1) + q(p_2) \to q(p_3) + q(p_4) + \gamma(k_1)$ have the following form:

$$\hat{s} = (p_1 + p_2)^2 = (k_1 + p_3 + p_4)^2, \quad \hat{t} = (p_1 - k_1)^2 = (p_4 + p_3 - p_2)^2,$$
$$\hat{u} = (p_2 - k_1)^2 = (p_4 + p_3 - p_1)^2, \quad \hat{q}_1 = (p_1 - p_3)^2 = (k_1 + p_4 - p_2)^2,$$
$$\hat{q}_2 = (p_2 - p_4)^2 = (k_1 + p_3 - p_1)^2, \quad \hat{s}_1 = (p_4 + p_3)^2 = (p_1 + p_2 - k_1)^2,$$
$$\hat{t}_1 = (p_1 - p_4)^2 = (k_1 + p_1 - p_2)^2, \quad \hat{u}_1 = (p_2 - p_3)^2 = (k_1 + p_4 - p_1)^2,$$
$$\hat{q}_3 = (p_3 + k_1)^2 = (p_1 + p_2 - p_4)^2, \quad \hat{q}_4 = (p_4 + k_1)^2 = (p_1 + p_2 - p_3)^2.$$

As independent invariants for 2 → 3 we can use invariants such as: $\hat{s}_1$, $\hat{q}_4$, $\hat{q}_1$, $\hat{u}$, $\hat{s}$. These five invariants are independent, and all the others can be expressed in terms of them and the particle masses.

The matrix elements for the subprocess $qq \to qq\gamma$ are as following:

$$M_1 = -\frac{4e^3 e_q \delta_{Col1Col3}\delta_{Col2Col4}\bar{U}(p_3)\gamma U(p_1)\bar{U}(p_4)\gamma(\hat{p}_2 - \hat{k}_1 + m_q)\hat{\varepsilon}^*(k_1)U(p_2)}{81(k_1 - p_2 + p_4)^2\left((k_1 - p_2)^2 - m_q^2\right)},$$

$$M_2 = -\frac{ee_q g_s^2 T^{Glu6}_{Col3Col1} T^{Glu6}_{Col4Col2} \bar{U}(p_3)\gamma U(p_1)\bar{U}(p_4)\gamma(\hat{p}_2-\hat{k}_1+m_q)\hat{\varepsilon}^*(k_1)U(p_2)}{9(k_1-p_2+p_4)^2((k_1-p_2)^2-m_q^2)},$$

$$M_3 = -\frac{4e^3 e_q \delta_{Col1Col3}\delta_{Col2Col4} \bar{U}(p_3)\gamma U(p_1)\bar{U}(p_4)\hat{\varepsilon}^*(k_1)(\hat{k}_1+\hat{p}_4+m_q)\gamma U(p_2)}{81(k_1-p_2+p_4)^2((-k_1-p_4)^2-m_q^2)},$$

$$M_4 = -\frac{ee_q g_s^2 T^{Glu6}_{Col3Col1} T^{Glu6}_{Col4Col2} \bar{U}(p_3)\gamma U(p_1)\bar{U}(p_4)\hat{\varepsilon}^*(k_1)(\hat{k}_1+\hat{p}_4+m_q)\gamma U(p_2)}{9(k_1-p_2+p_4)^2((-k_1-p_4)^2-m_q^2)},$$

$$M_5 = \frac{4e^3 e_q \delta_{Col1Col4}\delta_{Col2Col3} \bar{U}(p_3)\gamma U(p_2)\bar{U}(p_4)\gamma(-\hat{p}_2+\hat{p}_3+\hat{p}_4+m_q)\hat{\varepsilon}^*(k_1)U(p_1)}{81(p_2+p_3)^2((p_2-p_3-p_4)^2-m_q^2)},$$

$$M_6 = \frac{ee_q g_s^2 T^{Glu6}_{Col4Col1} T^{Glu6}_{Col3Col2} \bar{U}(p_3)\gamma U(p_2)\bar{U}(p_4)\gamma(-\hat{p}_2+\hat{p}_3+\hat{p}_4+m_q)\hat{\varepsilon}^*(k_1)U(p_1)}{9(p_2+p_3)^2((p_2-p_3-p_4)^2-m_q^2)},$$

$$M_7 = \frac{4e^3 e_q \delta_{Col1Col4}\delta_{Col2Col3} \bar{U}(p_3)\gamma U(p_2)\bar{U}(p_4)\hat{\varepsilon}^*(k_1)(\hat{k}_1+\hat{p}_4+m_q)\gamma U(p_1)}{81(p_3-p_2)^2((-k_1-p_4)^2-m_q^2)},$$

$$M_8 = \frac{ee_q g_s^2 T^{Glu6}_{Col4Col1} T^{Glu6}_{Col3Col2} \bar{U}(p_3)\gamma U(p_2)\bar{U}(p_4)\hat{\varepsilon}^*(k_1)(\hat{k}_1+\hat{p}_4+m_q)\gamma U(p_1)}{9(p_3-p_2)^2((-k_1-p_4)^2-m_q^2)},$$

$$M_9 = \frac{4e^3 e_q \delta_{Col1Col4}\delta_{Col2Col3} \bar{U}(p_4)\gamma U(p_1)\bar{U}(p_3)\gamma(\hat{p}_2-\hat{k}_1+m_q)\hat{\varepsilon}^*(k_1)U(p_2)}{81(k_1-p_2+p_3)^2((k_1-p_2)^2-m_q^2)},$$

$$M_{10} = \frac{ee_q g_s^2 T^{Glu6}_{Col4Col1} T^{Glu6}_{Col3Col2} \bar{U}(p_4)\gamma U(p_1)\bar{U}(p_3)\gamma(\hat{p}_2-\hat{k}_1+m_q)\hat{\varepsilon}^*(k_1)U(p_2)}{9(k_1-p_2+p_3)^2((k_1-p_2)^2-m_q^2)},$$

$$M_{11} = \frac{4e^3 e_q \delta_{Col1Col4}\delta_{Col2Col3} \bar{U}(p_4)\gamma U(p_1)\bar{U}(p_3)\hat{\varepsilon}^*(k_1)(\hat{k}_1+\hat{p}_3+m_q)\gamma U(p_2)}{81(k_1-p_2+p_3)^2((-k_1-p_3)^2-m_q^2)},$$

$$M_{12} = \frac{ee_q g_s^2 T^{Glu6}_{Col4Col1} T^{Glu6}_{Col3Col2} \bar{U}(p_4)\gamma U(p_1)\bar{U}(p_3)\hat{\varepsilon}^*(k_1)(\hat{k}_1+\hat{p}_3+m_q)\gamma U(p_2)}{9(k_1-p_2+p_3)^2((-k_1-p_3)^2-m_q^2)},$$

$$M_{13} = -\frac{4e^3 e_q \delta_{Col1Col3}\delta_{Col2Col4} \bar{U}(p_4)\gamma U(p_2)\bar{U}(p_3)\gamma(-\hat{p}_2+\hat{p}_3+\hat{p}_4+m_q)\hat{\varepsilon}^*(k_1)U(p_1)}{81(p_2-p_4)^2((p_2-p_3-p_4)^2-m_q^2)},$$

$$M_{14} = -\frac{ee_q g_s^2 T^{Glu6}_{Col3Col1} T^{Glu6}_{Col4Col2} \bar{U}(p_4)\gamma U(p_2)\bar{U}(p_3)\gamma(-\hat{p}_2+\hat{p}_3+\hat{p}_4+m_q)\hat{\varepsilon}^*(k_1)U(p_2)}{9(p_2-p_4)^2((p_2-p_3-p_4)^2-m_q^2)},$$

$$M_{15} = -\frac{4e^3 e_q \delta_{Col1Col3}\delta_{Col2Col4} \bar{U}(p_4)\gamma U(p_2)\bar{U}(p_3)\hat{\varepsilon}^*(k_1)(\hat{k}_1+\hat{p}_3+m_q)\gamma U(p_1)}{81(p_4-p_2)^2((-k_1-p_3)^2-m_q^2)},$$

$$M_{16} = -\frac{ee_q g T^{Glu6}_{Col3Col1} T^{Glu6}_{Col4Col2} \bar{U}(p_4)\gamma U(p_2)\bar{U}(p_3)\hat{\varepsilon}^*(k_1)(\hat{k}_1+\hat{p}_3+m_q)\gamma U(p_1)}{9(p_4-p_2)^2((-k_1-p_3)^2-m_q^2)}.$$

where *e*, $e_q$, $\alpha_E = \frac{e^2}{4\pi}$ and $\alpha_s = \frac{g_s^2}{4\pi} = \frac{4\pi}{\beta_0 \, ln\left(\frac{Q^2}{\Lambda^2}\right)}$ charge of electron, quark, fine-structure constant and strong coupling constant in QCD, correspondingly, $\beta_0$ is the Beta function first order (leading order, LO) in QCD, $\beta_0 = 11 - \frac{2}{3}n_f$, $n_f$ – number of active quarks (flavors) $Q$ is the transferred momentum, $\Lambda$ is the scale parameter in QCD.